\documentclass[
aps,
prx,
showpacs
amsmath,
amssymb,
preprint,
onecolumn,
floatfix,
longbibliography,
lengthcheck,
superscriptaddress
]{revtex4-1}

\usepackage{graphicx}
\usepackage{mathtools}
\usepackage{braket}
\usepackage{hyperref}

\begin{document}

\title{Ab-initio Optimized Effective Potentials for Real Molecules in Optical Cavities: Photon Contributions to the Molecular Ground state} 

\author{Johannes Flick}
  \email[Electronic address:\;]{johannes.flick@mpsd.mpg.de}
  \affiliation{Max Planck Institute for the Structure and Dynamics of Matter and Center for Free-Electron Laser Science \& Department of Physics, Luruper Chaussee 149, 22761 Hamburg, Germany}
\author{Christian Sch\"afer}
  \email[Electronic address:\;]{christian.schaefer@mpsd.mpg.de}
  \affiliation{Max Planck Institute for the Structure and Dynamics of Matter and Center for Free-Electron Laser Science \& Department of Physics, Luruper Chaussee 149, 22761 Hamburg, Germany}
  \author{Michael Ruggenthaler}
  \email[Electronic address:\;]{michael.ruggenthaler@mpsd.mpg.de}
  \affiliation{Max Planck Institute for the Structure and Dynamics of Matter and Center for Free-Electron Laser Science \& Department of Physics, Luruper Chaussee 149, 22761 Hamburg, Germany}
  \author{Heiko Appel}
  \email[Electronic address:\;]{heiko.appel@mpsd.mpg.de}
  \affiliation{Max Planck Institute for the Structure and Dynamics of Matter and Center for Free-Electron Laser Science \& Department of Physics, Luruper Chaussee 149, 22761 Hamburg, Germany}
  \author{Angel Rubio}
  \email[Electronic address:\;]{angel.rubio@mpsd.mpg.de}
  \affiliation{Max Planck Institute for the Structure and Dynamics of Matter and Center for Free-Electron Laser Science \& Department of Physics, Luruper Chaussee 149, 22761 Hamburg, Germany}

\date{\today}

\begin{abstract}
We introduce a simple scheme to efficiently compute photon exchange-correlation contributions due to the coupling to transversal photons as formulated in the newly developed quantum-electrodynamical density functional theory (QEDFT)~\cite{ruggenthaler2014,flick2015,ruggenthaler2015,ruggenthaler2011b,tokatly2013}. Our construction employs the optimized-effective potential (OEP) approach by means of the Sternheimer equation to avoid the explicit calculation of unoccupied states. We demonstrate the efficiency of the scheme by applying it to an exactly solvable GaAs quantum ring model system, a single azulene molecule, and chains of sodium dimers, all located in optical cavities and described in full real space. While the first example is a two-dimensional system and allows to benchmark the employed approximations, the latter two examples demonstrate that the correlated electron-photon interaction appreciably distorts the ground-state electronic structure of a real molecule. By using this scheme, we not only construct typical electronic observables, such as the electronic ground-state density, but also illustrate how photon observables, such as the photon number, and mixed electron-photon observables, e.g. electron-photon correlation functions, become accessible in a DFT framework. This work constitutes the first three-dimensional ab-initio calculation within the new QEDFT formalism and thus opens up a new computational route for the ab-initio study of correlated electron-photon systems in quantum cavities.
\end{abstract}

\date{\today}

\maketitle

\section{Introduction}
\noindent Over the past decades, methods in computational material science and quantum chemistry have been successfully applied to accurately model materials properties. Such material properties usually depend on the electronic structure of the system of interest that is dictated by the laws of quantum mechanics. Recently it has been demonstrated that by hybridizing light strongly with the electronic structure of the system, novel effects appear providing a promising route for a new design of material properties. Such recent experiments include, matter-photon coupling for living systems~\cite{coles2017}, vibrational strong coupling~\cite{george2016}, changes in chemical reactivity~\cite{thomas2016}, symmetry protected collisions of strongly interacting photons~\cite{thompson2017}{, the Bose-Einstein condensation~\cite{kasprzak2006} or the room-temperature polariton lasing~\cite{cohen2010} of exciton-polaritons, and ultra-strong coupling in circuit-QED~\cite{murch2016}} to mention a few. {Condensed matter systems driven out of equilibrium provide optional possibilities for novel properties, e.g. the creation of Floquet-Bloch states~\cite{wang2013} and Floquet-Weyl semimetals~\cite{huebener2017}. Additionally, the Floquet-scheme enables the development of new time-dependent DFT functionals with explicit memory dependence.} Recently, we and our co-workers have introduced a novel density-functional approach (QEDFT) to describe such complex dynamics of strongly interacting electrons, photons and phonon systems~\cite{ruggenthaler2011b,tokatly2013,ruggenthaler2014,pellegrini2015,flick2015, ruggenthaler2015}, all on the same theoretical footing. The framework of QEDFT is the first attempt to deal with the electron-photon interaction from first principles and has been demonstrated for the first time in Refs.~\cite{tokatly2013,ruggenthaler2014,flick2015,flick2016}. Together with new experiments on chemical systems in optical cavities~\cite{hutchison2012,thomas2016,george2016,ebbesen2016}, this work now opens up the field of Quantum-electrodynamical (QED) Chemistry and QED Materials~\cite{thomas2016,flick2017,ruggenthaler2017}. In this new field, so far model Hamiltonian schemes have also been used to successfully describe experiments~\cite{galego2015,galego2016}, but for an ab-initio description a full real-space picture is necessary.\\
As in any density-functional theory, the practical applicability of QEDFT is build upon the underlying approximations for the exchange-correlation (xc) functional. In contrast to traditional density-functional theory~\cite{kohn1965}, where a whole range of different approximation schemes for the xc functional are available~\cite{marques2012}, QEDFT still lacks a practical method to construct such approximations. {Previous works~\cite{tokatly2013, ruggenthaler2014, pellegrini2015} has opened the path to the development of such exchange-correlation functionals. Different routes are possible, e.g. functionals based on e.g. the electron density, the electronic orbitals or the electron current~\cite{schaefer2018} ultimately leading to the first quantitative accurate semilocal QEDFT functional.}
To close the gap, in this work, we introduce a simple, yet accurate, computational scheme to calculate the ab-initio xc potential for electronic systems coupled to quantized photon modes {based on the occupied electronic orbitals}. This method is based on the optimized effective potential (OEP) approach introduced by some of us in Ref.~\cite{pellegrini2015}. OEP has been previously used for purely electronic systems in DFT~\cite{kuemmel2003,kuemmel2008, engel2011}. In Ref.~\cite{pellegrini2015}, the construction of the OEP functional relies on the calculation of occupied and unoccupied orbitals. In particular the calculation of unoccupied states is computationally very demanding due to the large configuration space in any realistic many-body simulation and therefore hampers the practicability of the scheme. Here, we introduce a scheme that overcomes this limitation and does not involve the calculation of any unoccupied orbital. As a consequence, we find our scheme to be numerically highly efficient and thus we are able, for the first time, to calculate realistic molecular systems interacting with quantized photon modes from first principles. To achieve this goal, we employ the Sternheimer scheme~\cite{sternheimer1954} that allows us to construct the electron-photon OEP equation in an efficient manner. In this way, we only require the calculation of occupied orbitals together with solving linear equations which makes this approach computationally superior to the previous formulation. As a consequence our proposed scheme fits within the definition of a Kohn-Sham (KS) DFT as proposed by Axel Becke~\cite{becke2014} that defines KS-DFT as occupied-orbital-only. {Similar schemes have been used in the context of density-functional perturbation theory~\cite{baroni2001} and in many body-perturbation theory using the $GW$ self-energy approach~\cite{giustino2010}.}
\\
This paper is structured as follows: in section II, we introduce the formal framework to construct the ground-state xc potential using the OEP scheme. In section III, we apply the scheme to three different numerical examples and demonstrate the accuracy and the numerical feasibility for large-scale calculations. In the first example, we employ a model system for a GaAs quantum ring~\cite{rasanen2007, flick2015}. For this example, which is also exactly numerically solvable, we assess the accuracy of the scheme. We identify limiting cases, when to expect reliable results from the approximation. In the second example, we apply our method to a three-dimensional system, the azulene molecule in full three-dimensional real space. We demonstrate the effects of the correlated electron-photon interaction on the ground-state density. Additionally, we construct a mixed electron-photon correlation function that illustrates necessary ingredients for novel correlated electron-photon spectroscopies.  The last example of this paper treats chains of sodium dimers that allow us to systematically study the effects of many molecules in optical cavities. The latter two examples are the first QEDFT calculation for realistic molecules. { All these calculations demonstrate the reliability and applicability of the proposed numerical scheme. With realistic systems now calculatable, a promising avenue in the design of QED materials is introduced. }

\section{Theory}
We start by introducing the general nonrelativistic setup of the correlated electron-photon systems considered in the present work following previous works~\cite{tokatly2013,ruggenthaler2014,pellegrini2015,flick2016}. Let us consider $N_e=\sum_{i=\uparrow,\downarrow}N_\sigma=N_\uparrow+N_\downarrow$ interacting electrons of spin $\uparrow$ or $\downarrow$ located in an optical cavity. The electrons are coupled to $N_p$ quantized electromagnetic modes, i.e. photon modes. Each photon mode is identified by its cavity frequency $\omega_\alpha$ and polarization direction $\textbf{e}_\alpha$. We describe the matter-photon coupling in the Coulomb gauge, dipole approximation (long-wavelength approximation) and the length gauge~\cite{craig1998, tokatly2013}. The hereby emerging electron-photon coupling strength parameter $\lambda_\alpha$ is projected on the photon polarization direction $\boldsymbol \lambda_\alpha = \textbf{e}_\alpha \lambda_\alpha$. While in Coulomb gauge, the matter-photon interaction is explicitly described by the transversal degrees of freedom, the longitudinal degree of freedom leads to the Coulomb potential that describes the two-particle electron-electron interaction $1/|\textbf{r}_i-\textbf{r}_j|$. In this setup, the total electron-photon Hamiltonian reads~\cite{tokatly2013, pellegrini2015}
\begin{align}
\label{eq:full-ham}
\hat{H} &= \sum_{i=1}^{N_e}\left[-\frac{1}{2}\vec{\nabla}_i^2 + v_\text{ext}(\textbf{r}_i)\right]+\frac{1}{2}\sum_{i=1}^{N_e}\sum_{j=1,j\neq i}^{N_e}\frac{1}{|\textbf{r}_i-\textbf{r}_j|}\\
&+\sum_{\alpha=1}^{N_p}\left[\omega_\alpha \hat{a}^\dagger_\alpha \hat{a}_\alpha + \frac{j^{(\alpha)}_\text{ext}}{\omega_\alpha}\hat{q}_\alpha +  \hat{H}^{(\alpha)}_{ep}\nonumber\right],
\end{align}
where each photon mode is associated with a bosonic creation and annihilation operator ($\hat{a}^\dagger_\alpha$, $\hat{a}_\alpha$) that creates and destroys photons in the mode $\alpha$. The transversal electron-photon interaction $\hat{H}^{(\alpha)}_{ep}$ consists of two terms that read explicitly
\begin{align}
\hat{H}^{(\alpha)}_{ep} = -\sqrt{\frac{\omega_\alpha}{2}}\left(\hat{a}^\dagger_\alpha + \hat{a}_\alpha \right) \left(\boldsymbol\lambda_\alpha \cdot \textbf{R}\right)+\left(\boldsymbol\lambda_\alpha \cdot \textbf{R}\right)^2/2.
\label{eq:H_ep}
\end{align}
In dipole approximation, the electron-photon coupling comprises the electron dipole operator $\textbf{R}= \textbf{R}_0 + \sum_{i=1}^{N_e}\textbf{r}_i$ and the photon displacement coordinate $\hat{q}_\alpha = \sqrt{\frac{1}{2\omega_\alpha}}\left(\hat{a}^\dagger_\alpha + \hat{a}_\alpha \right)$. The electronic coordinates $\textbf{r}_i$ are defined with respect to the center of charge of the system $\textbf{R}_0$. As has been outlined in earlier work, using the creation and annihilation operators, we can setup the photon displacement and photon momentum operators $\hat{q}_\alpha$ and $\hat{p}_\alpha$~\cite{flick2015}. Physically these two operators are directly connected to the electric displacement field and the magnetic field, if summed over all modes~\cite{flick2015,flick2017}. The electron-photon coupling strength is given by
\begin{align}
\label{eq:ab_initio_coupling}
\boldsymbol \lambda_\alpha= \sqrt{4 \pi} S_\alpha(\textbf{k}_\alpha\cdot \textbf{R})\textbf{e}_\alpha,
\end{align}
where $S_\alpha$ denotes the mode function, e.g. a sine-function for the case of a cubic cavity~\cite{ruggenthaler2014,pellegrini2015} and $\textbf{k}_\alpha$ the wave vector. The effect of the nuclei employing the frozen-nuclei approximation enters the electron-photon Hamiltonian of Eq.~\ref{eq:full-ham} via the external potential $v_\text{ext}(\textbf{r})$. The effect of a static permanent dipole moment due to the nuclear charge can be neglected, since the two terms of Eq.~\ref{eq:H_ep} compensate each other in that case. For nuclei effects beyond the frozen-nuclei approximation, we refer the reader to Ref.~\cite{flick2017}, where electrons, nuclei and photons are treated on the same quantized footing.\\
Comparing QEDFT to standard DFT, we note that in QEDFT we have two sets of internal variables, i.e. the electron density $n(\textbf{r})$ and the photon displacement variables $q_\alpha$. It can be shown~\cite{ruggenthaler2015} that these two internal variables are in an one-to-one correspondence to the the external variables $v_\text{ext}(\textbf{r})$ and $j_\text{ext}^{(\alpha)}$. Here $j_\text{ext}^{(\alpha)}$ corresponds to the first-order time-derivative of an external charge current at time zero projected on the mode $\alpha$, i.e., $\int \mathrm{d}^3 r \; S_\alpha(\textbf{k}_\alpha\cdot \textbf{r}) \textbf{e}_\alpha \cdot \partial_t \textbf{j}_\text{ext}(\textbf{r},t)$  at $t=0$~\cite{tokatly2013,ruggenthaler2014}. The reason for the appearance of the time-derivative is the length-gauge transformation that rewrites the coupling to the photons in terms of the displacement field instead of in terms of the vector potential~\cite{tokatly2013,ruggenthaler2014}. Since the displacement field corresponds to the electric field minus the polarization~\cite{flick2017, rokaj2017}, and the electric field is the time derivative of the vector potential, the conjugate external variable to $q_{\alpha}$ needs to contain a time-derivative as well. In this work, we choose  $j_\text{ext}^{(\alpha)}=0$, without loss of generality. For $j_\text{ext}^{(\alpha)}\neq0$, or $\textbf{R}_0\neq \textbf{0}$ we can find a unitary transformation~\cite{dimitrov2017} that eliminates corresponding terms in Eq.~\ref{eq:full-ham}. {This transformation introduces instead a static external displacement field.} By exploiting the one-to-one correspondence of QEDFT, we find that all observables (electronic, photonic and mixed) become functionals of the internal variables. Formulated differently, any change in the internal variables will lead to changes in experimentally accessible observables.\\
In this work, we use the KS scheme~\cite{kohn1965} of density-functional theory introduced for electron-photon problems in Refs.~\cite{ruggenthaler2014,tokatly2013,flick2015} and commonly used in all DFT calculations. The KS scheme allows us to describe interacting many-body problems by the following set of effective noninteracting equations~\cite{tokatly2013}
\begin{align}
\hat{h}_{s\sigma}\varphi_{i\sigma}(\textbf{r}) = \Bigg[-\frac{1}{2}\vec{\nabla}_{i}^2 + v_{s\sigma}(\textbf{r})\Bigg]\varphi_{i\sigma}(\textbf{r}) = \epsilon_{i\sigma}\varphi_{i\sigma}(\textbf{r}).
\end{align}
for $N_\sigma$ Kohn-Sham orbitals $\varphi_{i\sigma}(\textbf{r})$ with spin $\sigma$. The effective Kohn-Sham potential $v_{s\sigma}(\textbf{r})$ is given by
\begin{align}
v_{s\sigma}(\textbf{r})=v_\text{ext}(\textbf{r})+v_{\text{Hxc}\sigma}(\textbf{r})+\sum^{N_p}_{\alpha=1}v^{(\alpha)}_{\text{Mxc}\sigma}(\textbf{r})
\label{eq:ks_vs}
\end{align}
and can be divided into the external potential $v_\text{ext}(\textbf{r})$, the usual Hartree-exchange-correlation (Hxc) potential $v_{\text{Hxc}\sigma}(\textbf{r})$ that accounts for the electron-electron interaction and the mode-dependent meanfield-exchange-correlation potential (Mxc) $v^{(\alpha)}_{\text{Mxc}\sigma}(\textbf{r})$~\footnote{In general electron-photon systems, we find that contributions due to the kinetic energy can not be attributed solely to the electron-electron or electron-photon interaction.}. Both, Hxc and Mxc contain the unknown exchange-correlation parts that have to be approximated. In exact KS-QEDFT, these parts are chosen such that the electron density $n(\textbf{r})$ that is the sum of the spin-resolved electron densities $n_\sigma(\textbf{r}) = \sum_{i=1}^{N_\sigma}\varphi^*_{i\sigma}(\textbf{r})\varphi_{i\sigma}(\textbf{r})$ is equivalent in the interacting and the noninteracting system. In the ground state, we have a simple connection between the exchange-correlation energy 
\begin{align}
E_{xc} = E^{(ee)}_{xc}+\sum^{N_p}_{\alpha=1} E^{(\alpha)}_{xc}
\label{eq:total-ex}
\end{align}
{that includes contributions from the electron-electron interaction $(ee)$ and the electron photon interaction $(\alpha)$} and the corresponding xc potential that reads as follows~\cite{engel2011}
\begin{align}
v_{xc\sigma}(\textbf{r}) = \frac{\delta E_{xc}}{\delta n_\sigma(\textbf{r})}.
\label{eq:vx-ex-dn}
\end{align}
This connection will be now exploited to setup the OEP equation. Throughout this work, we use the exchange-only approximation, i.e. $E_{xc} \approx E^{(ee)}_{x}+\sum^{N_p}_{\alpha=1} E^{(\alpha)}_{x}$. While we use the standard definition for $E^{(ee)}_{x}$~\cite{kuemmel2003,kuemmel2008}, i.e. the Fock energy, we focus in the following on the exchange energy due to the electron-photon interaction $E^{(\alpha)}_{x}$. The interaction terms in Eq.~\ref{eq:H_ep} contain the electron-photon coupling strength $\boldsymbol \lambda_\alpha$ in first-order and second-order. For the ground state the first-order exchange energy vanishes~\cite{pellegrini2015}, if the photons are not exposed to an external current $j_\text{ext}^{(\alpha)}$. Therefore, the leading order becomes the second-order in $\boldsymbol \lambda_\alpha$ and the exchange energy can be written as an orbital functional as~\cite{pellegrini2015}
\begin{align}
&E^{(\alpha)}_{x}(\{\varphi_{i\sigma}\},\{\Phi_{{i\sigma},\alpha}^{(1)}\},\{\Phi_{{i\sigma},\alpha}^{(2)} \}) = \label{eq:exchange_energy}\\
&\sum_{\sigma=\uparrow,\downarrow}\sum_{i=1}^{N_\sigma}\sqrt{\frac{\omega_\alpha}{8}} \bra{\Phi_{{i\sigma},\alpha}^{(1)}} \hat{d}_\alpha \ket{\varphi_{i\sigma}} +\frac{1}{4}  \bra{\Phi^{(2)}_{{i\sigma},\alpha}}\hat{d}_\alpha \ket{\varphi_{i\sigma}} + c.c.\nonumber
\end{align}
where $c.c.$ refers to the complex conjugate of all former terms. Additionally, we define the projected dipole operator $\hat{d}_\alpha = {\boldsymbol\lambda_\alpha} \cdot \textbf{r}$. As does the electron-photon interaction Hamiltonian in Eq.~\ref{eq:H_ep}, also the electron-photon exchange energy $E^{(\alpha)}_{x}$ consists of two parts, both containing different electronic orbital shifts. The first orbital shift is the solution of the following Sternheimer equation
\begin{align}
\left[\hat{h}_{s\sigma} - \left(\epsilon_{i\sigma} - \omega_\alpha\right)\right] {\Phi^{(1)}_{i\sigma,\alpha}}(\textbf{r}) &= -\sqrt{\frac{\omega_\alpha}{2}}\hat{d}_\alpha {\varphi_{i\sigma}}(\textbf{r}) \label{eq:1st-occ}\\
&+ \sqrt{\frac{\omega_\alpha}{2}}\sum_{k=1}^{N_\sigma}{d}^{(\alpha)}_{ki\sigma}{\varphi_{k\sigma}}\,(\textbf{r})\nonumber
\end{align}
with the matrix element $d^{(\alpha)}_{ij\sigma}=\bra{\varphi_{i\sigma}}\hat{d}_\alpha\ket{\varphi_{j\sigma}}$ and the orbital shift can be written explicitly as
\begin{align}
{\Phi^{(1)}_{i\sigma,\alpha}}(\textbf{r}) &= \sqrt{\frac{\omega_\alpha}{2}}\sum_{j=N_\sigma+1}^{\infty} 
\frac{{d}^{(\alpha)}_{ji\sigma}\;\varphi_{j\sigma}(\textbf{r})}{\epsilon_{i\sigma} - \epsilon_{j\sigma} - \omega_\alpha } \label{eq:1st-unocc}.
\end{align}
The second orbital shift is defined by
\begin{align}
\Phi^{(2)}_{i\sigma,\alpha}(\textbf{r}) &= \sum_{j=N_\sigma+1}^\infty d^{(\alpha)}_{ji\sigma}\; {\varphi_{j\sigma}}(\textbf{r})\label{eq:2nd-unocc}\\ 
&= \hat{d}_\alpha\varphi_{i\sigma}(\textbf{r}) - \sum_{k=1}^{N_\sigma}d^{(\alpha)}_{ki\sigma}\; \varphi_{k\sigma}(\textbf{r})\label{eq:2nd-occ}.
\end{align}
While both orbital shifts can be formulated explicitly in terms of all KS orbitals (in Eq.~\ref{eq:1st-unocc} and Eq.~\ref{eq:2nd-unocc}, respectively), only the second orbital shift ${\Phi^{(2)}_{i\sigma,\alpha}}$ can be formulated explicitly in terms of solely occupied orbitals given by Eq.~\ref{eq:2nd-occ}. However, the shift ${\Phi^{(1)}_{i\sigma,\alpha}}$ can be defined implicitly by a Sternheimer equation that only invokes occupied orbitals as given in Eq.~\ref{eq:1st-occ}.\\
Since the exchange energy given in Eq.~\ref{eq:exchange_energy} scales with $\lambda_\alpha^2$, the exchange energy is the Lamb shift of the ground state~\cite{pellegrini2015}. Thus all corrections for the ground state are in their magnitude on the order of the Lamb shift. For electron-photon problems, we we find that $E^{(\alpha)}_{x}$ as defined by Eq.~\ref{eq:exchange_energy} has a functional dependency on all occupied orbitals, and both orbital shifts. The standard route to obtain the OEP equation involves the calculation of functional derivatives of the orbitals and accordingly has to be generalized for the electron-photon case. In this case, we need consequently also the functional derivatives of the orbital shifts. Nevertheless, as will be demonstrated in the following, the standard route to construct the OEP equation can be adapted to accommodate this subtle difference. Having defined the total exchange energy $E_x$ in Eq.~\ref{eq:total-ex}, we now proceed to calculate the corresponding Kohn-Sham potential using functional derivatives. From Eq.~\ref{eq:vx-ex-dn}, we can setup the following OEP equation by using the chain rule of functional derivatives
\begin{widetext}
\begin{align}
\label{eq:vx-direct}
v_{x\sigma}(\textbf{r})=\sum_{\sigma',\sigma''=\uparrow,\downarrow}\sum_{\alpha=1}^{N_p}\sum_{i=1}^{N_\sigma}\iint d \textbf{r}'d \textbf{r}'' \frac{\delta v_{s\sigma'}(\textbf{r}')}{\delta n_{\sigma}(\textbf{r})}  \Bigg[&\frac{\delta \varphi_{i\sigma''}(\textbf{r}'')}{\delta v_{s\sigma'}(\textbf{r}')}\frac{\delta E_{x}}{\delta \varphi_{i\sigma''}(\textbf{r}'')}+ \frac{\delta \Phi^{(1)}_{i\sigma'',\alpha}(\textbf{r}'')}{\delta v_{s\sigma'}(\textbf{r}')}\frac{\delta E_{x}}{\delta \Phi^{(1)}_{i\sigma'',\alpha}(\textbf{r}'')}\\
&+\frac{\delta \Phi^{(2)}_{i\sigma'',\alpha}(\textbf{r}'')}{\delta v_{s\sigma'}(\textbf{r}')}\frac{\delta E_{x}}{\delta \Phi^{(2)}_{i\sigma'',\alpha}(\textbf{r}'')}\Bigg] + \text{c.c.}\nonumber
\end{align}
\end{widetext}
The OEP equation of Eq.~\ref{eq:vx-direct} contains several different terms that need an individual point-wise evaluation. First, we start discussing the functional derivatives of the exchange energy. These terms can be calculated straightforwardly using Eq.~\ref{eq:exchange_energy} and are given as follows~\footnote{Please note, that for brevity, we do not explicitly evaluate the $E^{ee}$ contributions, but state its implications if necessary.}
\begin{align}
\frac{\delta E_{x}}{\delta \varphi_{i\sigma}(\textbf{r})} &=    \sqrt{\frac{\omega_\alpha}{8}}\hat{d}_\alpha\Phi_{i\sigma,\alpha}^{*(1)}(\textbf{r})  +\frac{1}{4} \hat{d}_\alpha\Phi_{i\sigma,\alpha}^{*(2)}(\textbf{r}),   \\
\frac{\delta E_{x}}{\delta \Phi^{(1)}_{i\sigma,\alpha}(\textbf{r})}&= \sqrt{\frac{\omega_\alpha}{8}}\hat{d}_\alpha\varphi_{i\sigma}^*(\textbf{r}), \\
\frac{\delta E_{x}}{\delta \Phi^{(2)}_{i\sigma,\alpha}(\textbf{r})} &=
\frac{1}{4}\hat{d}_\alpha\varphi_{i\sigma}^*(\textbf{r}).
\end{align}
As the next step, we need to calculate the functional derivatives of the KS orbitals and orbital shifts with respect to the Kohn-Sham potential $v_s$. In the case of the KS orbitals, this derivative is given by~\cite{kuemmel2008,engel2011}
\begin{align}
\frac{\delta \varphi_{i\sigma}(\textbf{r}')}{\delta v_{s\sigma'}(\textbf{r})} &= \delta_{\sigma\sigma'}\sum_{j\neq i} \frac{\varphi_{j\sigma}^*(\textbf{r})\varphi_{i\sigma}(\textbf{r})}{\epsilon_{i\sigma}-\epsilon_{j\sigma}}\varphi_{j\sigma}(\textbf{r}'),
\label{eq:wavefunction-functional}
\end{align}
where the sum runs over all orbitals, except $i=j$. All remaining terms in Eq.~\ref{eq:vx-direct} are functional derivatives of the orbital shifts. We start by discussing $\Phi^{(2)}_{i\sigma,\alpha}(\textbf{r})$, since it is conceptually simpler to obtain, than $\Phi_{i\sigma,\alpha}^{(1)}(\textbf{r})$. From Eq.~\ref{eq:2nd-occ}, for an infinitesimal change in $\Phi^{(2)}_{i\sigma,\alpha}(\textbf{r})$, i.e. $\delta \Phi^{(2)}_{i\sigma,\alpha}(\textbf{r})$, by keeping only first-order terms and combining with Eq.~\ref{eq:wavefunction-functional}, we obtain
\begin{align}
&\frac{\delta \Phi^{(2)}_{i\sigma,\alpha}(\textbf{r}')}{\delta v_{s\sigma'}(\textbf{r})} = \delta_{\sigma\sigma'}\\
&\times\Bigg\{\sum_{j\neq i }\frac{\varphi_{j\sigma}^*(\textbf{r})\varphi_{i\sigma}(\textbf{r})}{\epsilon_{i\sigma}-\epsilon_{j\sigma}}\Bigg[\hat{d}_\alpha\;\varphi_{j\sigma}(\textbf{r}')-\sum_{k=1}^{N_\sigma}{d}^{(\alpha)}_{kj\sigma}\;\varphi_{k\sigma}(\textbf{r}')\Bigg]\nonumber\\
&-\sum_{k=1}^{N_\sigma}\sum_{j\neq k}\Bigg[\frac{\varphi_{j\sigma}^*(\textbf{r})\varphi_{k\sigma}(\textbf{r})}{\epsilon_{k\sigma}-\epsilon_{j\sigma}}{d}^{(\alpha)}_{ki\sigma}\;\varphi_{j\sigma}(\textbf{r}')\nonumber\\
&-\frac{\varphi_{k\sigma}^*(\textbf{r})\varphi_{j\sigma}(\textbf{r})}{\epsilon_{k\sigma}-\epsilon_{j\sigma}}{d}^{(\alpha)}_{ji\sigma}\;\varphi_{k\sigma}(\textbf{r}')\Bigg]\Bigg\}.\nonumber
\end{align}
The derivation of the remaining functional derivative of the first orbital shift, i.e. ${\delta \Phi^{(1)}_{i\sigma,\alpha}(\textbf{r})}/{\delta v_s(\textbf{r}')}$ is given in full detail in appendix \ref{sec:derivation_second_orbital_shift} and we only state the final result here
\begin{align}
\frac{\delta \Phi^{(1)}_{i\sigma,\alpha}(\textbf{r}')}{\delta v_{s\sigma'}(\textbf{r})} &=\delta_{\sigma\sigma'}\label{eq:second-oder-derivative}\\
\times\sqrt{\frac{\omega_\alpha}{2}}\sum_{j=N_\sigma+1}^{\infty}&\Bigg[\sum_{l=N_\sigma+1}^{\infty} \frac{{\varphi_{l\sigma}^*}(\textbf{r}) {\varphi_{j\sigma}}(\textbf{r})}{\epsilon_{i\sigma} - \epsilon_{l\sigma}-\omega_\alpha }\frac{{d}^{(\alpha)}_{ji\sigma}\;\varphi_{l\sigma}(\textbf{r}')}{\epsilon_{i\sigma}-\epsilon_{j\sigma}-\omega_\alpha}\nonumber\\
&- \frac{{\varphi_{i\sigma}^*}(\textbf{r}){\varphi_{i\sigma}}(\textbf{r})}{\epsilon_{i\sigma} - \epsilon_{j\sigma}-\omega_\alpha }\frac{{d}^{(\alpha)}_{ji\sigma}\;\varphi_{j\sigma}(\textbf{r}')}{\epsilon_{i\sigma}-\epsilon_{j\sigma}-\omega_\alpha}\nonumber\\
&+\sum_{l\neq i}   \frac{{\varphi_{l\sigma}^*}(\textbf{r}){\varphi_{i\sigma}}(\textbf{r})}{\epsilon_{i\sigma}-\epsilon_{l\sigma}} \frac{d^{(\alpha)}_{jl\sigma}\;\varphi_{j\sigma}(\textbf{r}')}{\epsilon_{i\sigma} - \epsilon_{j\sigma}-\omega_\alpha}\nonumber\\
&-\sum_{k=1}^{N_\sigma} \frac{{\varphi_{j\sigma}^*}(\textbf{r}){\varphi_{k\sigma}}(\textbf{r})}{\epsilon_{k\sigma}-\epsilon_{j\sigma}}\frac{ {d}^{(\alpha)}_{ki\sigma}\;\varphi_{j\sigma}(\textbf{r}')}{\epsilon_{i\sigma}-\epsilon_{j\sigma}-\omega_\alpha} \nonumber\\
&-\sum_{k=1}^{N_\sigma} \frac{{\varphi_{k\sigma}^*}(\textbf{r}){\varphi_{j\sigma}}(\textbf{r})}{\epsilon_{k\sigma}-\epsilon_{j\sigma}}\frac{ {d}^{(\alpha)}_{ji\sigma}\;\varphi_{k\sigma}(\textbf{r}')}{\epsilon_{i\sigma}-\epsilon_{j\sigma}-\omega_\alpha} \Bigg].\nonumber
\end{align}
Combining all these terms brings us to an alternative formulation of the OEP equation. By now plugging all ingredients into Eq.~\ref{eq:vx-direct} an alternative OEP equation can be derived that is given by the simple equation
\begin{align}
\label{eq:oep_equation}
&\sum_{i=1}^{N_\sigma}\int d\textbf{r}'M_{i\sigma}^*(\textbf{r}')G_{\text{S}i\sigma}(\textbf{r}',\textbf{r})\varphi_{i\sigma}(\textbf{r})-\Lambda_{i\sigma}(\textbf{r}) + c.c. = 0,
\end{align}
with the Kohn-Sham Green's function~\cite{kuemmel2008}
\begin{align}
G_{\text{S}i\sigma}(\textbf{r}',\textbf{r}) = \sum_{j\neq i}\frac{\varphi_{j\sigma}^*(\textbf{r})\varphi_{j\sigma}(\textbf{r}')}{\epsilon_{i\sigma}-\epsilon_{j\sigma}}.
\end{align}
Due to the energy dependence of $E^{(\alpha)}_x$, we find that the nonvanishing additional inhomogeneity~\cite{engel2011} $\Lambda_{i\sigma}(\textbf{r})$ is given by
\begin{align}
\Lambda_{i\sigma}(\textbf{r}) = \frac{1}{2}\sum^{N_p}_{\alpha=1}&\Biggl[{\Phi}^{(1)*}_{i\sigma,\alpha}(\textbf{r}){\Phi}^{(1)}_{i\sigma,\alpha}(\textbf{r})\\
&- \braket{{\Phi}^{(1)}_{i\sigma,\alpha}|{\Phi}^{(1)}_{i\sigma,\alpha}}\varphi^{*}_{i\sigma}(\textbf{r})\varphi_{i\sigma}(\textbf{r}) \Biggr]\nonumber
\end{align}
and the orbital shift $M_{i\sigma}(\textbf{r})$ by
\begin{align}\label{eq:soep3}
M_{i\sigma}^*(\textbf{r}) =& -\left(v_{x\sigma}(\textbf{r}) - u_{xi\sigma}(\textbf{r})\right)\varphi^*_{i\sigma}(\textbf{r})\\
&+\sum_{\alpha=1}^{N_p}\Biggl[\hat{d}_\alpha\Bigg(\sqrt{\frac{\omega_\alpha}{2}}\Phi^{(1)*}_{i\sigma,\alpha}(\textbf{r})+\frac{1}{2}\hat{d}_\alpha \varphi^*_{i\sigma}(\textbf{r})\Bigg)
\nonumber\\
&-\sum_{k=1}^{N_\sigma}{d}^{(\alpha)}_{ik\sigma}\Bigg(\sqrt{\frac{\omega_\alpha}{2}}{\Phi}^{(1)*}_{k\sigma,\alpha}(\textbf{r}) + \;\hat{d}_{\alpha} \varphi^*_{k\sigma}(\textbf{r})\Bigg)\Biggr].\nonumber
\end{align}
The orbital shift $M_{i\sigma}(\textbf{r})$ contains in the first line the electron-electron interaction, we choose the exchange-only approximation, i.e.
\begin{align}
u_{xi\sigma}(\textbf{r})=\frac{1}{\varphi^*_{i\sigma}(\textbf{r})}\frac{\delta E^{(ee)}_{x}[\{\varphi_{j\tau}\}]}{\delta \varphi_{i\sigma}(\textbf{r})},
\end{align}
and $ E^{(ee)}_{x}$ is the usual Fock exchange energy. The following lines are corrections due to the correlated electron-photon interaction that induce density changes in the electronic system~\cite{pellegrini2015}.\\
The main advantage of the present reformulation is that we can write the OEP equation for electron-photon problems in a simple form. {This formulation is similar to Refs.~\cite{kuemmel2003, engel2011} that provide the formulation for electrons-only.}
\begin{align}\label{eq:soep}
\sum_{i=1}^{N_\sigma}\psi^*_{i\sigma}(\textbf{r})\varphi_{i\sigma}(\textbf{r}) - \Lambda_{i\sigma}(\textbf{r}) + c.c. = 0.
\end{align}
and the orbital shifts $\psi^*_{i\sigma}(\textbf{r})$ can be obtained using a Sternheimer equation
\begin{align}\label{eq:soep2}
\left(\hat{h}_{s\sigma} - \epsilon_{i\sigma}\right)\psi^*_{i\sigma}(\textbf{r}) = M_{i\sigma}^*(\textbf{r}) - \braket{M_{i\sigma}|\varphi_{i\sigma}}\varphi^*_{i\sigma}(\textbf{r})
\end{align}
This equation has to be solved self-consistently with Eq.~\ref{eq:1st-occ}. By this procedure, we have replaced the problem of calculating the OEP equation using all unoccupied states by a problem of solving $N_p$+1 Sternheimer equations that only invoke occupied orbitals.

\subsection{Novel Types of Observables}

One of the advantages of QEDFT over DFT is the correct treatment of the quantum nature of the photon field {and its interaction with a correlated many-body electron system}. Thus, by exploiting the one-to-one correspondence of the internal variables to the external variables~\cite{ruggenthaler2015}, the photon observables (and the electronic observables) become functionals of the internal variables. Therefore, if we know the internal variables and their functional dependency, we can construct arbitrary observables. In the case of orbital functionals, we can use the KS orbitals to construct these observables. In this section, we now introduce new types of observables into the DFT framework, i.e. photonic observables and observables of mixed matter-photon character. The first example for a photonic observable is the number of photons in each mode. This observable can be calculated in terms of KS orbital shifts as follows
\begin{align}
\label{eq:ada}
n^{pt}_\alpha = \langle \hat{a}^\dagger_\alpha \hat{a}_\alpha \rangle =& \sum_{\sigma=\uparrow,\downarrow}\sum_{i=1}^{N_\sigma}\langle{{\Phi}^{(1)}_{i\sigma,\alpha}}|{{\Phi}^{(1)}_{i\sigma,\alpha}}\rangle \\
&+  \frac{({\boldsymbol\lambda_\alpha} \cdot \textbf{R}_0)^2}{2\omega_\alpha}\nonumber.
\end{align}
Physically, we can attribute the orbital shifts that are calculated by Eq.~\ref{eq:2nd-occ} with wave-function corrections that carry each a single photon. Thus, we can use these shifts to calculate the photon number in each photon mode. While the first term in Eq.~\ref{eq:ada} is due to the quantum fluctuations of the photon mode, the latter term is a classical contribution due to a nonvanishing $\textbf{R}_0$. By performing this connection, we assume that the photon number is dominated by contributions stemming from single-photon processes. To this end, we can expect a good quality of this photon number observables if this is the case, while if many-photon processes contribute we expect poorer results.
\\
Examples for mixed electron-photon observables~\cite{ruggenthaler2017} are electron-photon correlation functions. For instance, the charge-density-displacement-field correlation function $A^{(\alpha)}(\textbf{r})$ {we define} as
\begin{align}
\label{eq:correlation-function}
A^{(\alpha)}(\textbf{r}) &= \bra{\Psi_0}\hat{n}(\textbf{r}) (\hat{a}_\alpha^\dagger + \hat{a}_\alpha)  \ket{\Psi_0} \\
&= \sum_{\sigma=\uparrow,\downarrow}\sum_{i=1}^{N_\sigma}{\varphi}_{i\sigma}(\textbf{r}){{\Phi}^{(1)}_{i\sigma,\alpha}}(\textbf{r}) + c.c.\\
&+\sqrt{\frac{2}{\omega_\alpha}}{(\boldsymbol \lambda_\alpha \cdot \textbf{R}_0)}n(\textbf{r})\nonumber,
\end{align}
where $\Psi_0$ denotes the many-body ground state of the system. The given {expression} is the leading term in {an expansion in orders of} $\boldsymbol\lambda_\alpha$. {Physically this correlation function corresponds to the local forces that the displacement field of the photons exerts on the electrons~\cite{tokatly2013, flick2015}. If we perturb the photon field, the change of these local forces will rearrange the charges in an intricate manner. While for a classical field $A^{(\alpha)}(\textbf{r})$ merely becomes a product of the (positive) electronic density and the value of the displacement field and is therefore only positive or negative, in the quantum case $A^{(\alpha)}(\textbf{r})$ can locally change its sign. Consequently probing this correlation function spectroscopically could allow to obtain novel insights into structural properties of complex systems.}

\subsection{Krieger-Li-Iafrate approximation}
\label{sec:KLI}
As will be demonstrated in the application section, the OEP equation leads to accurate results. However, since the xc potential is given only implicitly by Eq.~\ref{eq:oep-S} and Eq.~\ref{eq:oep-vxc-new}, it may be hard to converge. One way to circumvents this problem and to obtain an explicit formula for the xc potential is the Krieger-Li-Iafrate (KLI) approximation~\cite{krieger1990, krieger1992construction, krieger1992systematic}.\\
In contrast to the common Coulomb OEP equation~\cite{kuemmel2003}, in the case of correlated electron-photon coupling an additional inhomogeneous contribution appears, i.e. $\Lambda_{i\sigma}$.
The consequence of this structural deviation from the well known OEP equation in the electronic case, where no inhomogeneity is present complicates the common approximation schemes. A direct energy denominator approximation does not only leave an arbitrariness on the remaining energy denominator but the corresponding approximations leave divergent contributions uncanceled. The reformulation in terms of Sternheimer shifts avoids unbalanced approximations in divergent contributions.
If we multiply Eq.~\ref{eq:soep} with the Kohn-Sham potential~\cite{kuemmel2003} and decompose Eq.~\ref{eq:soep} and Eq.~\ref{eq:soep2} starting from
\begin{align}
	\begin{split}
		v_{s\sigma}(\textbf{r})\psi^*_{i\sigma}(\textbf{r}) &= -\left(-\frac{1}{2}\vec{\nabla}^2_{i}-\epsilon_{i\sigma}\right)\psi^*_{i\sigma}(\textbf{r})\\ &+ M_{i\sigma}^*(\textbf{r}) - \braket{M_{i\sigma}|\varphi_{i\sigma}}\varphi^*_{i\sigma}(\textbf{r})~,	
	\end{split}
\end{align}
with Eq.~\ref{eq:soep3} and
\begin{align}\label{eq:skli1}
	\begin{split}
		&\varphi_{i\sigma}(\textbf{r})\left(-\frac{1}{2}\vec{\nabla}^2_{i}-\epsilon_{i\sigma}\right)\psi_{i\sigma}^{*}(\textbf{r}) + v_{s\sigma}(\textbf{r}) \Lambda_{i\sigma}(\textbf{r})\\ &= \varphi_{i\sigma}(\textbf{r})(\hat{h}_{s\sigma}(\textbf{r}) - \varepsilon_{i\sigma}) \psi_{i\sigma}^{*}(\textbf{r}),
	\end{split}
\end{align}
we arrive at the exact reformulation
\begin{align}
	{v}_{x\sigma}( \textbf{r}) &= 
	\frac{1}{n_\sigma(\textbf{r})}\sum\limits_{i=1}^{N_{\sigma}} \notag \langle \varphi_{i\sigma}| v_{x\sigma} | \varphi_{i\sigma}\rangle \vert \varphi_{i\sigma}(\textbf{r})\vert^2\\ &+ \frac{1}{n_\sigma(\textbf{r})}\Re \sum\limits_{i=1}^{N_{\sigma}} \bigg[\big\{{M}^{*}_{i \sigma}(\textbf{r})+v_{x\sigma}(\textbf{r}) \varphi_{i\sigma}^{*}(\textbf{r})\big\} \notag\\ &- \langle \big\{{M}_{i \sigma}+v_{x\sigma} \varphi_{i\sigma}\big\} \vert \varphi_{i \sigma}\rangle \varphi_{i\sigma}^*(\textbf{r}) \\ & - (\hat{h}_{s\sigma}(\textbf{r}) - \varepsilon_{i\sigma}) \psi_{i\sigma}^{*}(\textbf{r}) \bigg] \varphi_{i\sigma}(\textbf{r})~. \notag
\end{align}
The common approximation scheme now assumes $(\hat{h}_{s\sigma}(\textbf{r}) - \varepsilon_{i\sigma}) \psi_{i\sigma}^{*}(\textbf{r}) = 0$ which is exact for a single electron if no inhomogeneity was present in  Eq.~\ref{eq:soep}. A corresponding substitution involving $ \psi_{i\sigma}^{*}(\textbf{r}) \approx  \Lambda_{i\sigma}(\textbf{r})/\varphi_{i\sigma}(\textbf{r}) $ leads in the general case to nodal points. The variety of possibilities result in different deficiencies and inconsistencies (see also Engel et al. \cite{engel1998}). To remain as consistent as possible we decide to assume $(\hat{h}_{s\sigma}(\textbf{r}) - \varepsilon_{i\sigma}) \psi_{i\sigma}^{*}(\textbf{r}) = 0$ and the KLI equation reads then
\begin{align}
	\label{eq:kli}
	{v}_{x\sigma}(\textbf{r}) &= 
	\frac{1}{n_\sigma(\textbf{r})}\sum\limits_{i=1}^{N_{\sigma}} \notag \langle \varphi_{i\sigma}| v_{x\sigma} | \varphi_{i\sigma}\rangle \vert \varphi_{i\sigma}(\textbf{r})\vert^2\\ &+ \frac{1}{n_\sigma(\textbf{r})}\Re \sum\limits_{i=1}^{N_{\sigma}} \bigg[\big\{{M}^{*}_{i \sigma}(\textbf{r})+v_{x\sigma}(\textbf{r}) \varphi_{i\sigma}^{*}(\textbf{r})\big\}\\ &- \langle \big\{{M}_{i \sigma}+v_{x\sigma} \varphi_{i\sigma}\big\} \vert \varphi_{i \sigma}\rangle \varphi_{i\sigma}^*(\textbf{r}) \bigg] \varphi_{i\sigma}(\textbf{r})~. \notag
\end{align}
This reformulation avoids the solution of Eq.~\ref{eq:soep2} and can be solved explicitly for the exchange potential as a linear equation. This improves the stability with respect to the initial guess and represents in many cases a valuable starting-point for the OEP calculation. The KLI effectively neglects off-diagonal contributions to the response function mediated by the exchange potential.
Connecting to this, the accuracy reduces with increasing local dipole-moment which will especially manifest in the overestimation in local density perturbation under cavity influence.

\subsection{Numerical details}

We have implemented the OEP equation of Eq.~\ref{eq:soep} and the corresponding KLI equation of Eq.~\ref{eq:kli} in the {OCTOPUS} package~\cite{marques2003,castro2006,andrade2014}. The OEP equation can be solved using standard algorithms as e.g. described in Ref.~\cite{kuemmel2003}, i.e. in a self-consistent field (SCF) cycle. To obtain the numerical algorithm, we reformulate Eq.~\ref{eq:soep} as follows
\begin{align}
S_\sigma(\textbf{r}) = \sum_{i=1}^{N_\sigma}\psi^*_{i\sigma}(\textbf{r})\varphi_{i\sigma}(\textbf{r}) - \Lambda_{i\sigma}(\textbf{r}) + c.c.
\label{eq:oep-S}
\end{align}
The quantity $S_\sigma(\textbf{r})$ becomes a measure for convergence, since it is vanishing in the case of convergence (compare Eq.~\ref{eq:oep-S} and Eq.~\ref{eq:soep}). To obtain the new potential in the SCF step, we use
\begin{align}
v_{x\sigma}^{(new)}(\textbf{r}) = v_{x\sigma}^{(old)}(\textbf{r}) + c(\textbf{r})S_\sigma(\textbf{r}).
\label{eq:oep-vxc-new}
\end{align}
Different schemes to calculate $c(\textbf{r})$ are possible~\cite{hollins2012}, e.g. dividing by the electron density~\cite{kuemmel2003b}, using the Barzilai-Borwein minimizer~\cite{barzilati1988}, or connecting to conjugate-gradient algorithms~\cite{hollins2012}. However for our purpose, we find that choosing a constant value~\cite{kuemmel2003} is already sufficient and already provides the most stable and reliable algorithm. Thus, we choose $c(\textbf{r})=0.1$ a.u. for the azulene molecule and $c(\textbf{r})=20$ a.u. for the sodium chains. \\
As in the case of electronic OEP~\cite{kuemmel2003,krieger1992systematic}, we also find for the photon OEP that we can add an 
arbitrary (spatial-independent) constant to the exchange potential that does not alter the physical results. If we follow the lines of the electronic OEP~\cite{kuemmel2003,krieger1992systematic} and enforce the condition $\bra{\varphi_{N_\sigma\sigma}}v_{x\sigma}\ket{\varphi_{N_\sigma\sigma}} = \bra{\varphi_{N_\sigma\sigma}}M_{i\sigma}\ket{\varphi_{N_\sigma\sigma}}$, we find that in the single electron case, the single Kohn-Sham energy deviates from the total energy. From a physical point-of-view it is desirable that both coincides to connect to ionization energies. We find by fixing $\bra{\varphi_{N_\sigma\sigma}}v_{x\sigma}\ket{\varphi_{N_\sigma\sigma}}$ to the contribution of the highest occupied orbital to the exchange energy defined in Eq.~\ref{eq:exchange_energy}, i.e.
\begin{align}
&\bra{\varphi_{N_\sigma\sigma}}v_{x\sigma}\ket{\varphi_{N_\sigma\sigma}} = \bra{\varphi_{N_\sigma\sigma}}u_{xN_\sigma\sigma}\ket{\varphi_{N_\sigma\sigma}} \\
&+ \sum^{N_p}_{\alpha=1}\sqrt{\frac{\omega_\alpha}{8}} \bra{\Phi_{{N_\sigma\sigma},\alpha}^{(1)}} \hat{d}_\alpha \ket{\varphi_{N_\sigma\sigma}} +\frac{1}{4}  \bra{\Phi^{(2)}_{{N_\sigma\sigma},\alpha}}\hat{d}_\alpha \ket{\varphi_{N_\sigma\sigma}}\nonumber
\end{align}
we can restore this connection. However, we note that for the electronic OEP~\cite{kuemmel2003,krieger1992systematic} both routes coincide. Furthermore since in the present study we focus on energy differences, the arbitrary constant only modifies the offset in the presented xc potentials. 

\section{Numerical application}

\begin{figure}[ht]
\centerline{\includegraphics[width=0.5\textwidth]{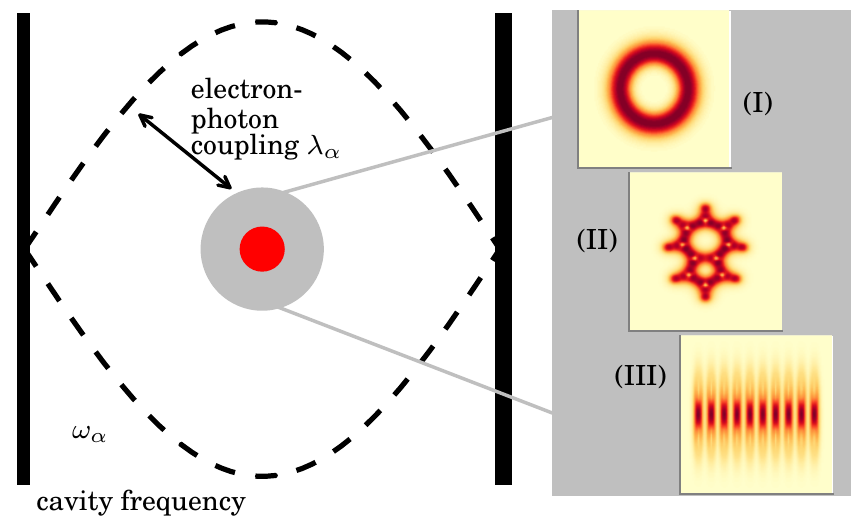}}
\caption{Overview of the three molecular systems in an optical cavity studied in the present work: (I) GaAs quantum ring, (II) single azulene molecule, (III) chain of ten Na$_2$ dimers all of which are coupled to a single photon mode with frequency $\omega_\alpha$ and electron-photon coupling strength $\lambda_\alpha$.}
\label{fig:model-system}
\end{figure}

As numerical applications, we analyze different examples in 2D and 3D. The first example is used to demonstrate the accuracy of the employed method. To this end, we benchmark the OEP scheme with an exactly solvable model system, i.e. a GaAs quantum ring located in an optical cavity~\cite{flick2015,flick2017b}, where we have published exact results previously~\cite{flick2015,flick2017b}. In this way we are able to validate the presented scheme before in the next examples, we apply it to real systems. Thus in the second example, we solve the electron-photon OEP equation for the first time in full three-dimensional real space. We study the azulene molecule and report the changes in the ground-state density if the molecule is located inside an optical cavity. The last example deals with realistic ensembles of molecules in optical cavities. Here we study the ground-state density of chains of sodium dimers with different length. The different examples studied in this work are schematically depicted in Fig.~\ref{fig:model-system}.

\subsection{GaAs quantum ring in an optical cavity}

We start by discussing the model for a GaAs semiconductor quantum ring coupled to a single photon mode in an optical cavity. The model consists of a single electron restricted to two spatial dimensions in real-space ($\textbf{r}=r_x\textbf{e}_x + r_y\textbf{e}_y$) interacting with the single photon mode with frequency $\hbar\omega_\alpha=1.41$~meV and polarization direction $\textbf{e}_\alpha=(1,1)$. The polarization direction enters via the electron-photon coupling strength, i.e. ${\boldsymbol \lambda_\alpha}=\lambda_\alpha \textbf{e}_\alpha$ and depends on the specific experimental setup. We choose the photon mode frequency in resonance with the first electronic transition. The external potential of the single electron is given by the following formula
\begin{align}
v_\text{ext}(\textbf{r}) = \frac{1}{2}m_0 \omega_0^2 \textbf{r}^2 + V_0 e^{-\textbf{r}^2/{d^2}} {,}
\label{eq:external-sym}
\end{align}
with parameters $\hbar\omega_0=10$~meV, $V_0=200$~meV, $d=10$~nm, {$m_0=0.067m_e$}~\cite{rasanen2007,flick2017b}. For the electron-photon coupling strength, we choose two values, i.e. in weak coupling with $\lambda_\alpha=0.0034$  meV$^{1/2}$/nm and in strong coupling $\lambda_\alpha=0.1342$  meV$^{1/2}$/nm. The effective three-dimensionality of this problem (two-dimensional electron and one-dimensional photon mode) is low enough that an exact solution is still accessible via exact diagonalization~\cite{flick2014}. To obtain the exact ground state, we employ a two-dimensional grid of $N=127$ grid points in each direction with a grid spacing of $\Delta x=0.7052$~nm to describe the single electron. The photon field is represented in the photon number eigenbasis and we include up to $41$ Fock number states. Using the exact wave function, we can numerically construct the exact exchange-correlation potential~\cite{flick2015,helbig2009}.
\begin{figure}[ht]
\centerline{\includegraphics[width=0.5\textwidth]{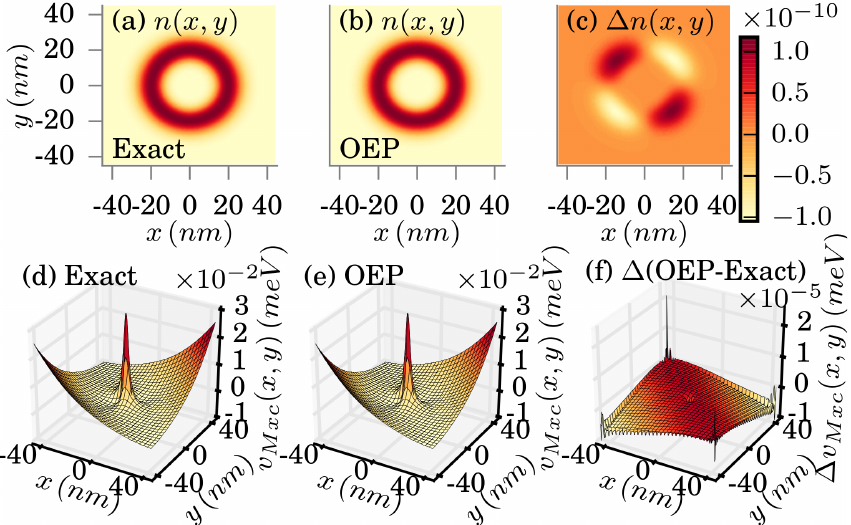}}
\caption{Exact (a) and OEP approximated (b) electron density in the weak-coupling limit ($\lambda_\alpha=0.0034$  meV$^{1/2}$/nm). The difference is shown in (c). The corresponding xc potentials are shown in (d) and (e), respectively. (f) shows the difference in the xc potentials.}
\label{fig:oep-rs-gaas-weak}
\end{figure}
We start by discussing the weak-coupling limit, where $\lambda_\alpha=0.0034$  meV$^{1/2}$/nm. In Fig.~\ref{fig:oep-rs-gaas-weak} (a), we show the exact ground-state density obtained by exact diagonalization. Compared to the bare electronic ground-state (for $\lambda_\alpha=0$) that also has a ring structure in the weak-coupling limit, we find small distortions~\cite{flick2017b}. In Fig.~\ref{fig:oep-rs-gaas-weak} (b), we show the OEP ground-state density and in (c) the difference between the exact and the OEP ground-state density. The deviation of the OEP ground-state density to the exact ground-state density is very small and in the order of magnitude of $10^{-10}$, i.e. close to our numerical precision. This high precision of the approximate electron density has its origin in the high quality of the OEP approximation for the xc potentials. The exact and the OEP xc potential are plotted in (d) and (e), respectively. In (f) we plot the difference of the exact to the OEP potential and find significant differences ($\mathcal{O}\sim 10^{-5}$) only in low-density regions, i.e. at the border of our grid. In contrast the inner high-density regions are well approximated leading to the accurate description of the electron density. This larger error can also be attributed to the algorithm, since low density regions are harder to converge in the OEP scheme. However, since low density regions do not contribute much to observables such as the total energy, this error will effectively not influence the overall result.
\begin{figure}[ht]
\centerline{\includegraphics[width=0.5\textwidth]{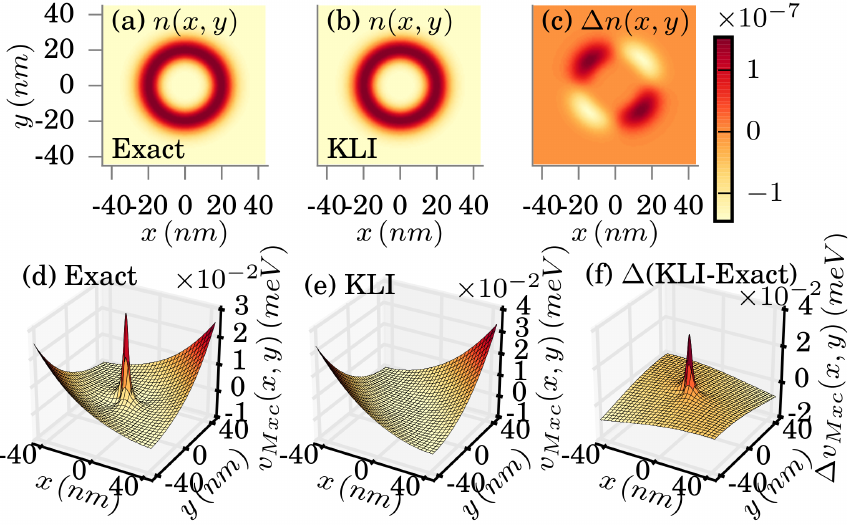}}
\caption{Exact (a) and KLI approximated (b) electron density in the weak-coupling limit ($\lambda_\alpha=0.0034$  meV$^{1/2}$/nm). The difference is shown in (c). The corresponding xc potentials are shown in (d) and (e), respectively. (f) shows the difference in the xc potentials. }
\label{fig:oep-rs-gaas-weak-kli}
\end{figure}
In Fig.~\ref{fig:oep-rs-gaas-weak-kli} we show how the KLI approximation (Sec.~\ref{sec:KLI}) performs in the weak-coupling limit for the single-electron case. In (b) we plot the KLI approximated electron density and in (c), we show the difference to the exact reference. We find errors in the electron density in the order of $\mathcal{O}\sim 10^{-7}$ that are due to the KLI xc potential. The KLI xc potential is shown in (e). We find that in comparison to the exact reference shown in (d) the overall shape of the potential is approximated correctly, while the peak in the middle of the potential is missing. The deviations can be also seen in (f), where we plot the difference between the exact and the KLI xc potential.
\begin{table}
\begin{center}
\begin{tabular}{ l | cccc  }
theory&pot&$\lambda_\alpha$ [(meV$^{1/2}$/nm)]&$E_{tot}$ [meV]&$n_{pt}$\\\hline
exact&s&0.0034&33.8782&0.0004738\\
OEP&s&0.0034&33.8782&0.0004730\\
KLI&s&0.0034&33.8782&0.0004727\\\hline
exact&s&0.1342&35.3072&3.1926\\
OEP&s&0.1342&35.3349&3.4011\\
exact&a&0.1342&32.4816&2.2053\\
OEP&a&0.1342&32.4875&2.2087\\
\end{tabular}
\end{center}
\caption{Results of the self-consistent KS calculation for the GaAs quantum ring in an optical cavity: The total energy $E_{tot}$ and photon number  $n_{pt}$ for different levels of theory, coupling strength and symmetric (s) and asymmetric (a) potentials are shown in the table.}
\label{tab:GaAs}
\end{table}
To quantify the differences for this example, we print the results of our calculations in Tab.~\ref{tab:GaAs}. The first three rows show the exact, OEP and KLI results for the total energy $E_{tot}$ and the photon number $n_{pt}$ in the weak-coupling limit using the external potential of Eq.~\ref{eq:external-sym}. Overall, we find a very good performance, of the OEP and KLI approximations. The OEP performs slightly better, but also the KLI gives accurate energies and photon numbers.
\begin{figure}[ht]
\centerline{\includegraphics[width=0.5\textwidth]{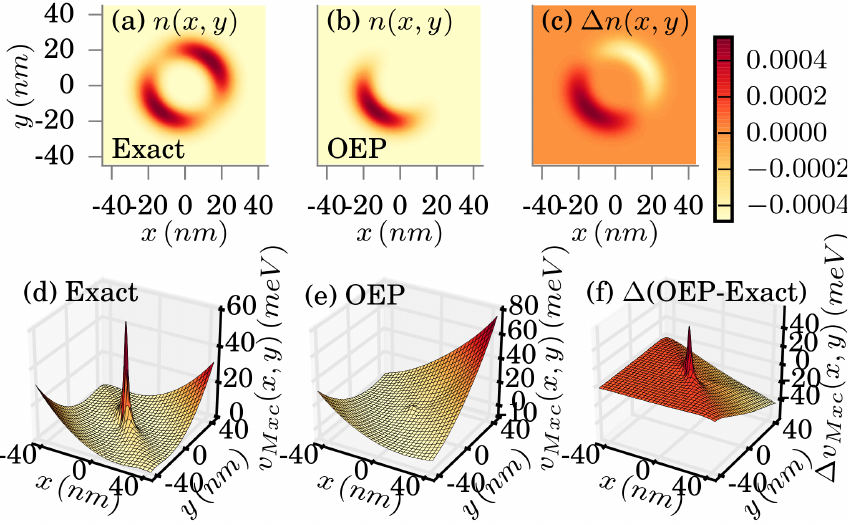}}
\caption{Exact (a) and OEP approximated (b) electron density in the strong-coupling limit ($\lambda_\alpha=0.1342$ meV$^{1/2}$/nm). The difference is shown in (c). The corresponding xc potentials are shown in (d) and (e), respectively. (f) shows the difference in the xc potentials.}
\label{fig:oep-rs-gaas-strong}
\end{figure}
Let us now analyze the strong-coupling limit. In Fig.~\ref{fig:oep-rs-gaas-strong}, we show the results obtained in the strong-coupling regime ($\lambda_\alpha=0.1342$ meV$^{1/2}$/nm), where we find a deviation in the exact ground-state electron density from the ring structure in the weak-coupling regime to a double-well structure~\cite{flick2017} as shown in Fig.~\ref{fig:oep-rs-gaas-strong} (a). This splitting is accompanied by a higher peak in the xc potential in the center of the grid as shown in Fig.~\ref{fig:oep-rs-gaas-strong} (d). Although in the weak-coupling limit, we find a very high accuracy of the OEP approximation, in the strong coupling limit, we observe the break-down of the OEP approximation. In Fig.~\ref{fig:oep-rs-gaas-strong} (b), we find that the OEP predicts an electron density that is located in only one of the two subwells with a wrong xc potential shown in Fig.~\ref{fig:oep-rs-gaas-strong} (e). Consequently the error of the OEP density and the potential shown in Fig.~\ref{fig:oep-rs-gaas-strong} (c) and (f) are very high.
\begin{figure}[ht]
\centerline{\includegraphics[width=0.5\textwidth]{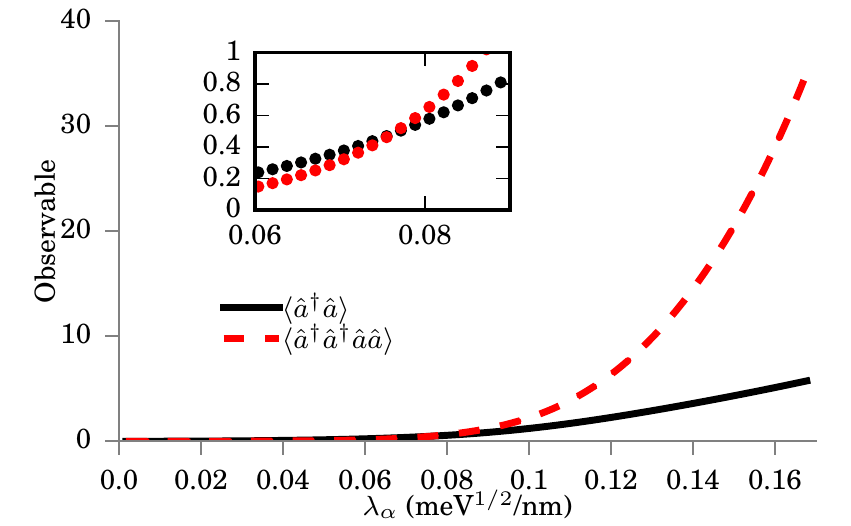}}
\caption{Plot of the photon number occupation $\langle \hat{a}^\dagger\hat{a}\rangle$ and double photon number $\langle \hat{a}^\dagger\hat{a}^\dagger\hat{a}\hat{a}\rangle$ for the GaAs quantum ring as function of the electron-photon coupling strength $\lambda_\alpha$. The inset shows the region $\lambda_\alpha \in \left[0.06,0.09 \right]$ meV$^{1/2}$/nm. When the double occupancy becomes significant, the OEP approximation start to fail (see text for more detail).}
\label{fig:oep-rs-gaas-adadaa-ada}
\end{figure}
The origins of this failure of the OEP can be understood by calculating the photon number $\langle \hat{a}^\dagger\hat{a}\rangle$ and the double occupancy $\langle \hat{a}^\dagger\hat{a}^\dagger\hat{a}\hat{a}\rangle$ in the photon mode shown in Fig.~\ref{fig:oep-rs-gaas-adadaa-ada}. Scaling the electron-photon coupling strength $\lambda_\alpha$ from the weak to the strong coupling limit, we find {that} two-photon processes become the dominant contributions to the ground state, when the electron density {splits}~\cite{flick2017b}. Since the OEP approximation by construction only considers single photon processes, its failure in this region is a natural consequence of the higher weight of two (and more) photon processes in the setting of the xc potential. In Ref.~\cite{flick2017b}, we have calculated the exact eigenvalues and find a close degeneracy of the ground state and the first-excited state in the strong-coupling limit (reminiscent of static correlation in quantum chemistry~\cite{dimitrov2016}). Similarly as in the electron-only case, where static correlation can be described by including correlation effects beyond exact exchange, in correlated electron-photon problems, we conclude that in the strong coupling limit going beyond exact exchange, i.e. single photon processes, to higher order processes, i.e. two-photon, three-photon, etc. is required to accurately describe this limit.
\begin{figure}[ht]
\centerline{\includegraphics[width=0.5\textwidth]{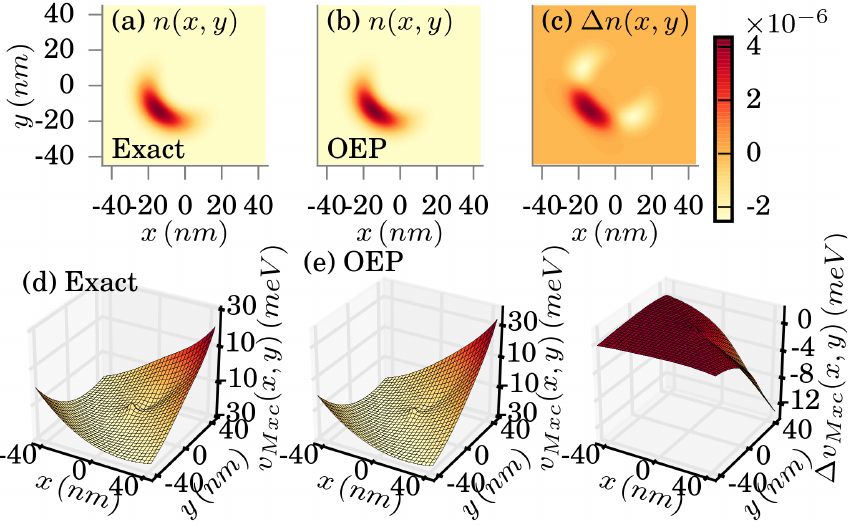}}
\caption{Exact (a) and OEP approximated (b) electron density in the strong-coupling limit ($\lambda_\alpha=0.1342$ meV$^{1/2}$/nm). The difference is shown in (c). The corresponding xc potentials are shown in (d) and (e), respectively. (f) shows the difference in the xc potentials.}
\label{fig:oep-rs-gaas-strong-asymmetric}
\end{figure}
In the last example, we study an asymmetric example in the strong-coupling limit ($\lambda_\alpha=0.1342$ meV$^{1/2}$/nm), where the external potential is given by
\begin{align}
\tilde{v}_\text{ext}(\textbf{r}) = \tilde{v}_\text{ext}(\textbf{r})+\bar{V}_0 \, \textbf{e}_\alpha\cdot \textbf{r}{.}
\end{align}
with $\bar{V}_0=0.1123$~meV/nm. The cavity frequency is again chosen to be in resonance to the first electronic excitation, i.e. $\hbar\omega_\alpha=2.72$~meV. The results are shown in Fig.~\ref{fig:oep-rs-gaas-strong-asymmetric}. We find while the density is approximated accurately with errors in the order of $10^{-6}$, also observables such as the photon number listed in Tab.~\ref{tab:GaAs} are approximated quite accurately due to the dominant mean-field contribution in Eq.~\ref{eq:ada}.\\
As conclusion, we have demonstrated in this section that the photon OEP is capable of describing a wide range of parameters correctly. In the weak-coupling regime, we have found highly accurate results. Additionally, we find in the strong coupling regime a failure of the OEP in the symmetric setup, while in the asymmetric setup, we have again an accurate description of the electron density. Having at hand a scheme to derive approximations for general functionals, we can also investigate novel types of observables that are not accessible with traditional DFT but need a QEDFT calculation. In the case at hand we find, for instance, good agreement for the photon number, where both the OEP and KLI approximation yield reliable results. Next, after we have assessed the quality of our approximations, we turn to real systems and show that QEDFT is an efficient ab-initio scheme to determine properties of complex systems coupled to photons.

\subsection{Azulene (C$_{10}$H$_8$) molecule in an optical cavity}
\begin{figure}[ht]
\centerline{\includegraphics[width=0.5\textwidth]{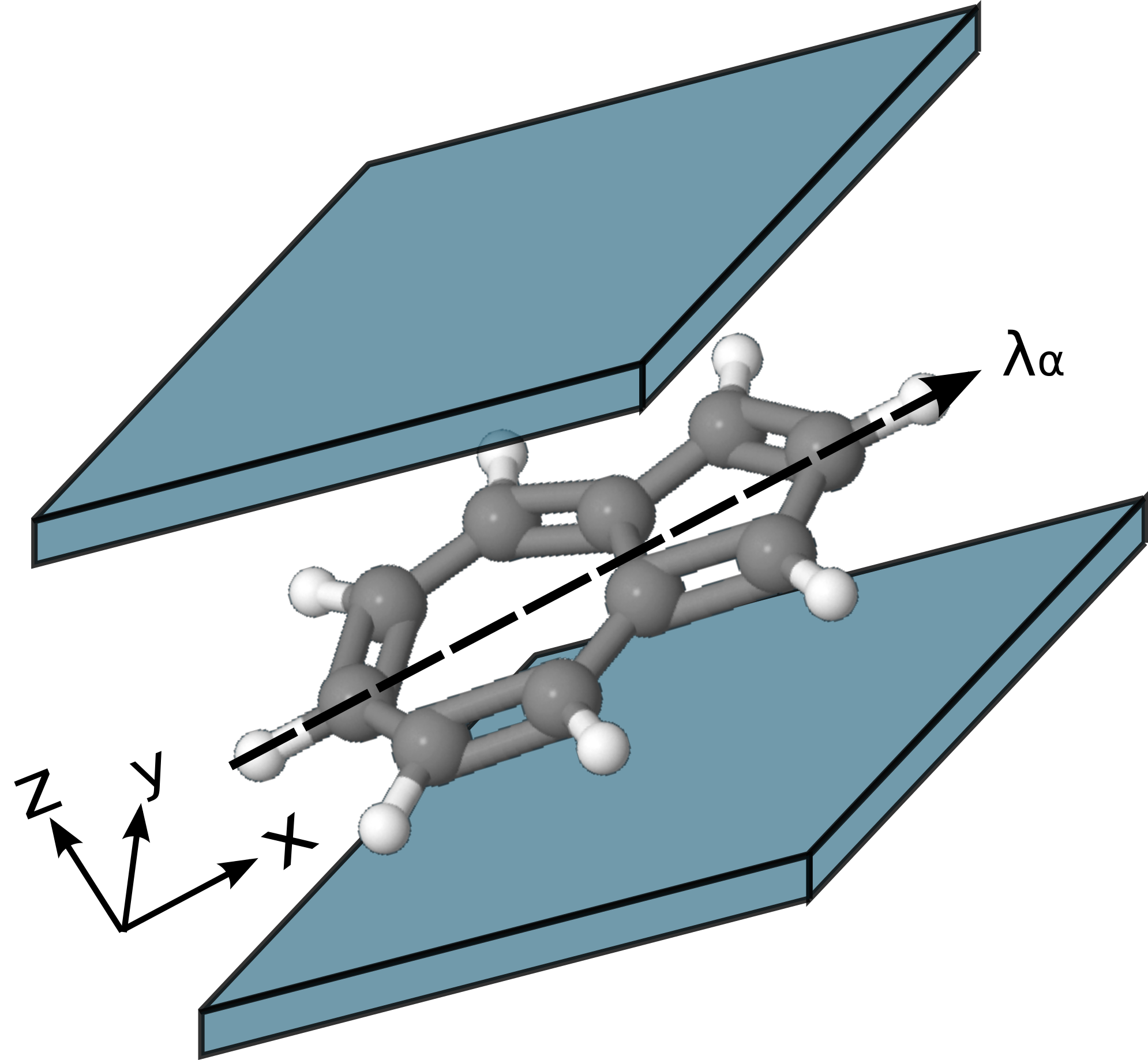}}
\caption{Azulene (C$_{10}$H$_8$) molecule in an optical cavity, ${\boldsymbol\lambda_\alpha}$ denotes the polarization direction of the photon field.}
\label{fig:azulene-molecule}
\end{figure}
\begin{table}
\begin{center}
\begin{tabular}{ l | ccccc  }
theory&24-1 [eV] &25-24 [eV]& $E_{tot}$ [eV]& $E^{ee}_x$ [eV]& $E^{ep}_x$ [eV]\\\hline
KLI& 16.57 & 2.24 & -1648.39 & -501.79 & 0.00 \\
OEP& 16.68 & 2.42 & -1648.53 & -503.04 & 0.00 \\
KLI-PT& 16.48 & 2.25 & -1644.38 & -502.11 & 3.89 \\
OEP-PT& 16.66 & 2.54 & -1644.71 & -503.67 & 3.79
\end{tabular}
\end{center}
\caption{Results of self-consistent KS calculation for real 3D azulene in an optical cavity: Energy difference between the highest occupied orbital (HOMO) (24th orbital) and the lowest occupied orbital (1st orbital), energy difference between the lowest unoccupied orbital (LUMO) (25th orbital) and the highest occupied orbital (HOMO), the total energy $E_{tot}$, the exchange energy $E_x^{ee}$ and photon exchange energy $E_x^{(\alpha)}$ for different levels of theory.}
\label{tab:azulene}
\end{table}
Our next example is the first real application of the QEDFT framework to a three-dimensional molecule, i.e the azulene (C$_{10}$H$_8$) molecule. To find a reliable equilibrium structure and determine the cavity frequency, we follow the following route. {First, we obtain the 3D conformer structure for azulene from the PubChem database~\cite{kim2015} (CID: 9231). Second, we use the geometry optimization of the OCTOPUS package employing the LDA functional~\cite{kohn1965,perdew1981} to calculate a relaxed ground-state structure. Third,} using this relaxed structure, we use the electron OEP to calculate a HOMO-LUMO gap of {$2.41$} eV that serves as the cavity frequency, i.e. {$\hbar\omega_\alpha = 2.41$}~eV. The electron-photon coupling includes the polarization direction of the photon field that is polarized along the x-direction with a strength of $\lambda_\alpha=0.08$, i.e. $\boldsymbol \lambda_\alpha = 0.08 \textbf{e}_x$~\footnote{For standard experimental parameters, e.g. for a single  trapped-atom cavity as described in  Ref.~\cite{khitrova2006} (Fig.~3), (V = $18,148$~$\mu$m$^3$), we deduce an experimental value of $\lambda_0=8,32 \times 10^{-10} a.u.$.}. In this example we want to investigate the question how the correlated electron-photon interaction alters the electronic ground-state density $n_0(\textbf{r})$. To numerically calculate the ground-state density of the azulene molecule, we use a grid of dimensions $16\times18\times8$ Bohr in $xyz$ directions. The grid spacing is chosen to be $\Delta x= 0.32$ \AA \; and to describe the core electrons of the carbon and hydrogen atoms we use {LDA} Troullier-Martins pseudopotentials~\cite{troullier1993}. Thus, we explicitly describe the 48 valence electrons in our calculations amounting to 24 doubly occupied Kohn-Sham orbitals. As described in the previous section, to describe the electron-electron interaction, we use the Fock exchange energy~\cite{kuemmel2003} also in the OEP setting.
\begin{figure}[ht]
\centerline{\includegraphics[width=0.5\textwidth]{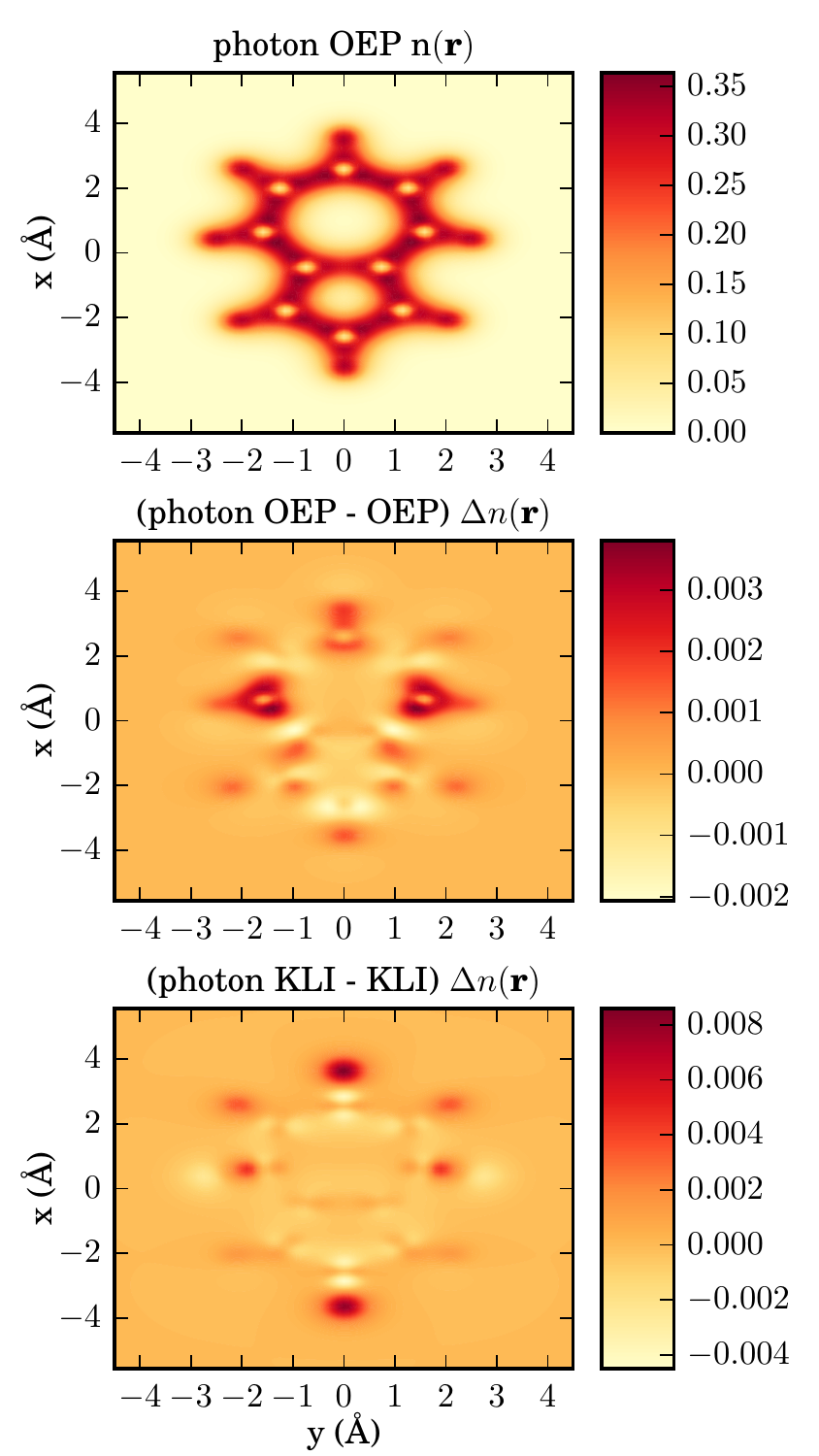}}
\caption{From top to bottom as a cut in x-y plane: OEP ground-state density of azulene, difference of electron-photon OEP and electron OEP ground-state density, and difference of electron-photon KLI and electron KLI ground-state density.}
\label{fig:oep-rs-azulene-xy}
\end{figure}
In Fig.~\ref{fig:oep-rs-azulene-xy}, we show in the top panel the ground-state density of the molecule in an optical cavity as a cut in the x-y plane. The electrons are highly localized in-between the nuclei. The aromatic ring structure induced by the arrangement of the carbon atoms is inherited in the ground-state electron density that has naturally the same symmetry. The middle plot of Fig.~\ref{fig:oep-rs-azulene-xy} shows the difference of the electronic ground-state density exposed to electron-photon coupling to the bare electronic ground-state density, i.e. the change in the density by going from gas phase to the case inside the cavity. The figure shows a rich fine structure in the center of the molecule, but also a pronounced accumulation region of electronic density at the top and bottom rim of the molecule. The plot on the bottom of Fig.~\ref{fig:oep-rs-azulene-xy} show the results of the KLI approximation. While the KLI seems to fail to correctly describe the rich inner structure of the density differences $\Delta n(\textbf{r})$, it correctly predicts the density accumulation regions at the top and bottom of the molecule. However, these regions are overestimated by a factor $\sim4$. To quantify the effects of the quantized electron-photon interaction on many-electron systems, we have provided numerical results in Tab.~\ref{tab:azulene}. For different levels of theory, we print the energy difference between lowest and highest occupied orbitals ($24-1$), the HOMO-LUMO gap ($25-24$), the total energy $E_{tot}$, and the electronic and the electron-photonic part of the exchange energy $E^{ee}_x$, and $E^{ep}_x$, respectively. For the given parameters, the electron-photon exchange energy is in the order of $\sim 3.8$eV and two orders of magnitude smaller than the electronic exchange energy $E^{ee}_x$ that is roughly $\sim 500$ eV. {As could be expected, the changes due to the coupling to the vacuum of the cavity are small in the ground state, i.e., we have determined the Lamb shift. However, due to the electron-photon coupling we now have access to novel types of observables.}
\begin{figure}[ht]
\centerline{\includegraphics[width=0.5\textwidth]{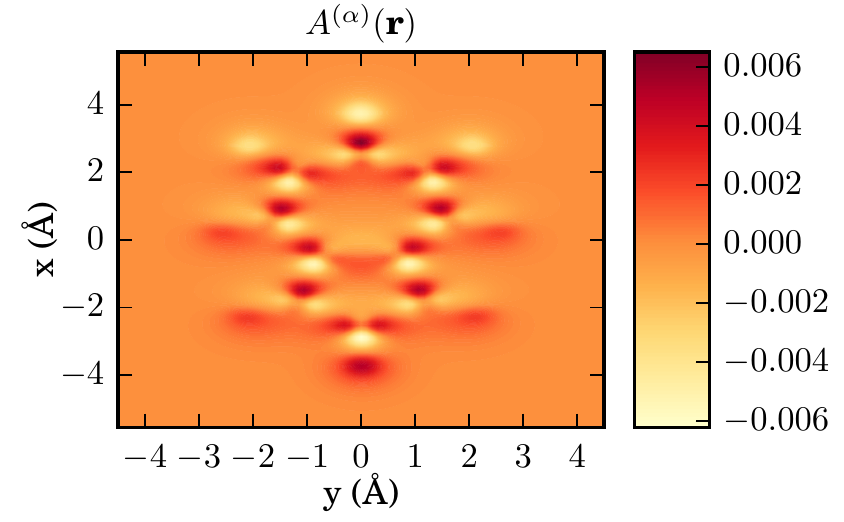}}
\caption{Correlation function as a cut in x-z plane $A^{(\alpha)}(\textbf{r})$ as defined in Eq.~\ref{eq:correlation-function}, calculated for the azulene molecule.}
\label{fig:oep-azulene-correlator}
\end{figure}
To connect to the novel mixed electron-photon observables within the framework of QEDFT, we calculate the correlator $A^{(\alpha)}(\textbf{r})$ as defined in Eq.~\ref{eq:correlation-function} without the mean-field contributions in Fig.~\ref{fig:oep-azulene-correlator}. {We find that the resulting local-force map due to the coupling to the photons indeed shows a complex structure with local sign changes. It indicates the forces that the electrons experience due the displacement field. Indeed, the local forces nicely agree with the rearrangement of the charge density upon coupling to the vacuum field. If we would perturb the photon mode, the electrons would experience forces in different directions depending on their position in the molecule. In contrast, a classical field in dipole approximation would only induce forces in one direction.} In conclusion, in this section we have presented the distorting effects of the quantized electron-photon interaction on molecules in cavities. We find that in QEDFT new observables become numerically {accessible} that {could} allow for novel experimental spectroscopic setups~\cite{ruggenthaler2017}.

\subsection{Chain of Sodium Dimers}
The last example that is studied in this paper {is} a chain of sodium dimers of variable length, i.e. up to ten sodium dimers. {We use this set up to highlight that QEDFT allows to investigate problems of quantum optics from first principles. For instance, we can consider the reliability of the ubiquitous Dicke model~\cite{garraway2011}, where many two-level systems are coupled to a cavity mode. This model predicts that due to the collective behavior of the two-level systems the usually weak coupling of the matter to the photon mode is effectively increased. This collective effect is one way of reaching the strong coupling limit in experiment. Still, due to the many simplifying assumptions employed in deriving this model some implications are debated, e.g., the superradiant phase transition~\cite{knight1978, vukics2014}. With a first-principles approach such as QEDFT many of these assumption can be avoided which could shed new light on these issues. Here we will not target these more intricate problems but rather show that we can recover from first principles the increase in the effective coupling strength. We do so by analyzing the behavior of the correlated electron-photon ground-state, when more and more emitters are coupled to the cavity field}\\
For this setup, we use the parameters for a sodium dimer given in Ref.~\cite{kuemmel2000}. For the optical cavity frequency, we choose the energy of the 3s-3p transition, i.e. $\omega_\alpha = 2.19$ eV. We assume that the photon field is polarized along the direction of the sodium chains with a strength of $\lambda_\alpha=0.006$, i.e. $\boldsymbol \lambda_\alpha = 0.006 \textbf{e}_y$. To calculate the chain of sodium dimers (Na$_2$), we use the sodium pseudopotentials and equilibrium distances for a single sodium dimer of Ref.~\cite{kuemmel2000}. For the real-space grid, we use dimensions $30\times \text{min}(30,N_c\times10) \times30$ Bohr with grid spacing $0.5$ Bohr, where $N_c$ is the chain length. The distance between the sodium dimers is chosen as $d=10$ Bohr.
\begin{figure}[ht]
\centerline{\includegraphics[width=0.5\textwidth]{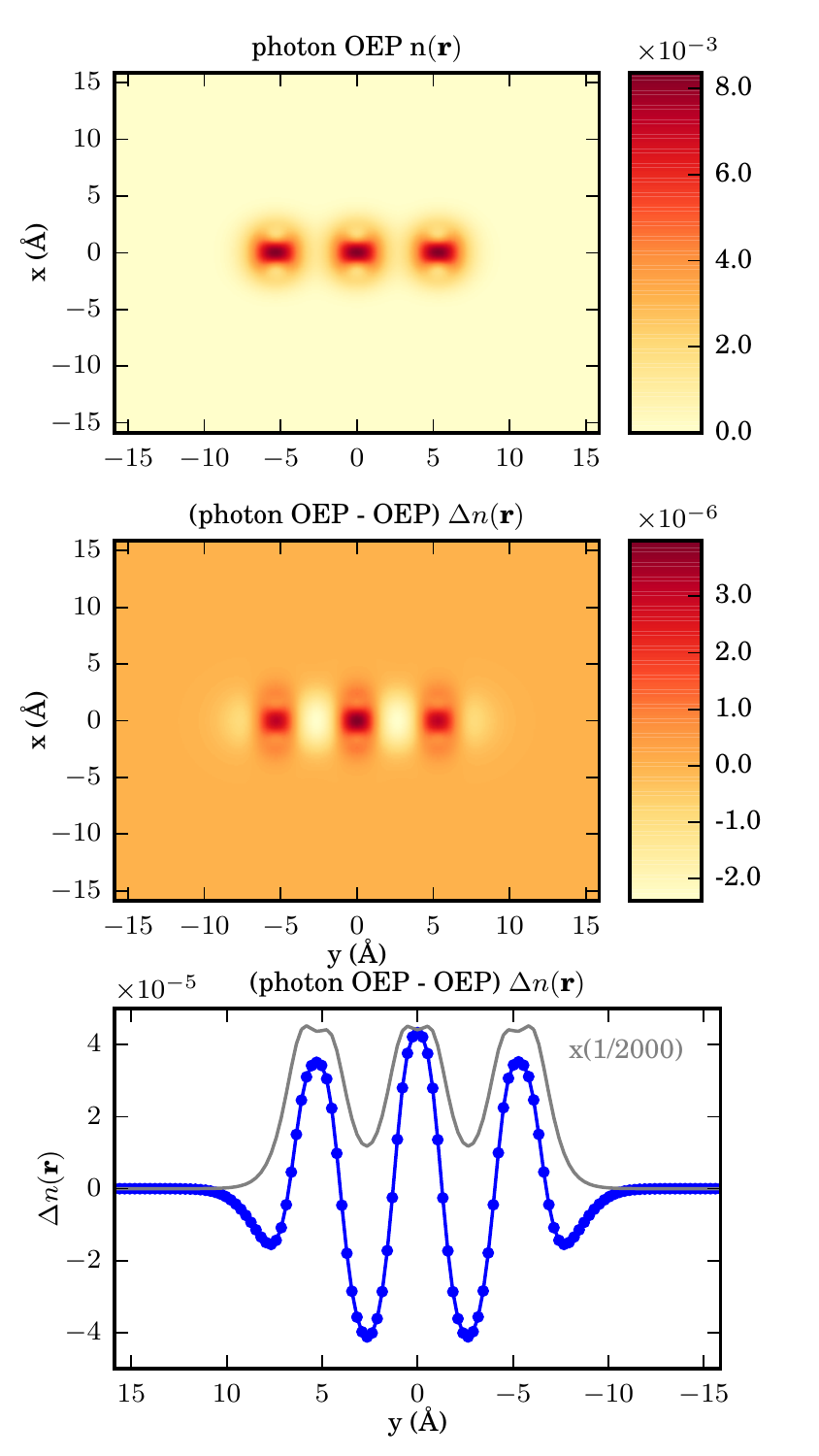}}
\caption{From top to bottom as a cut in x-y plane: OEP ground-state density of three sodium dimers, difference of electron-photon OEP and electron OEP ground-state density, and cut of difference of electron-photon OEP and electron OEP along the y axis {in blue against the electron-photon OEP density in grey. Please note that the latter has been reduced by a factor of $1/2000$}.}
\label{fig:oep-rs-sodium-3}
\end{figure}
The case of three sodium dimers is illustrated in Fig.~\ref{fig:oep-rs-sodium-3}. As in the previous example, in the top panel we show a cut of the ground-state electron density in the x-y plane. Each sodium dimer contains two electrons and the electrons are localized between the sodium nuclei. In the middle plot, we show the difference in the electron density of the system with and without the cavity. {The lower plot in Fig.~\ref{fig:oep-rs-sodium-3} shows a cut along the y-axis in blue against the ground-state electron density in the cavity in grey. We find three maxima for density accumulation and four minima from where the density has been rearranged. Further, we find that the electron-photon interaction pushes the electron density onto high-density regions. This density accumulation stems from regions in-between the dimers, where the amount of density is decreasing in the cavity.}\\
The next figure, Fig.~\ref{fig:oep-rs-sodium-10} shows the case of ten sodium dimers. The first plot shows the electron density of the ground state. In the second plot we see the difference of the electron density of the system inside the optical cavity to the bare electron density. Between the maxima, we find local minima where electron density is rearranged, as shown in the plot in the bottom.
\begin{figure}[ht] 
\centerline{\includegraphics[width=0.5\textwidth]{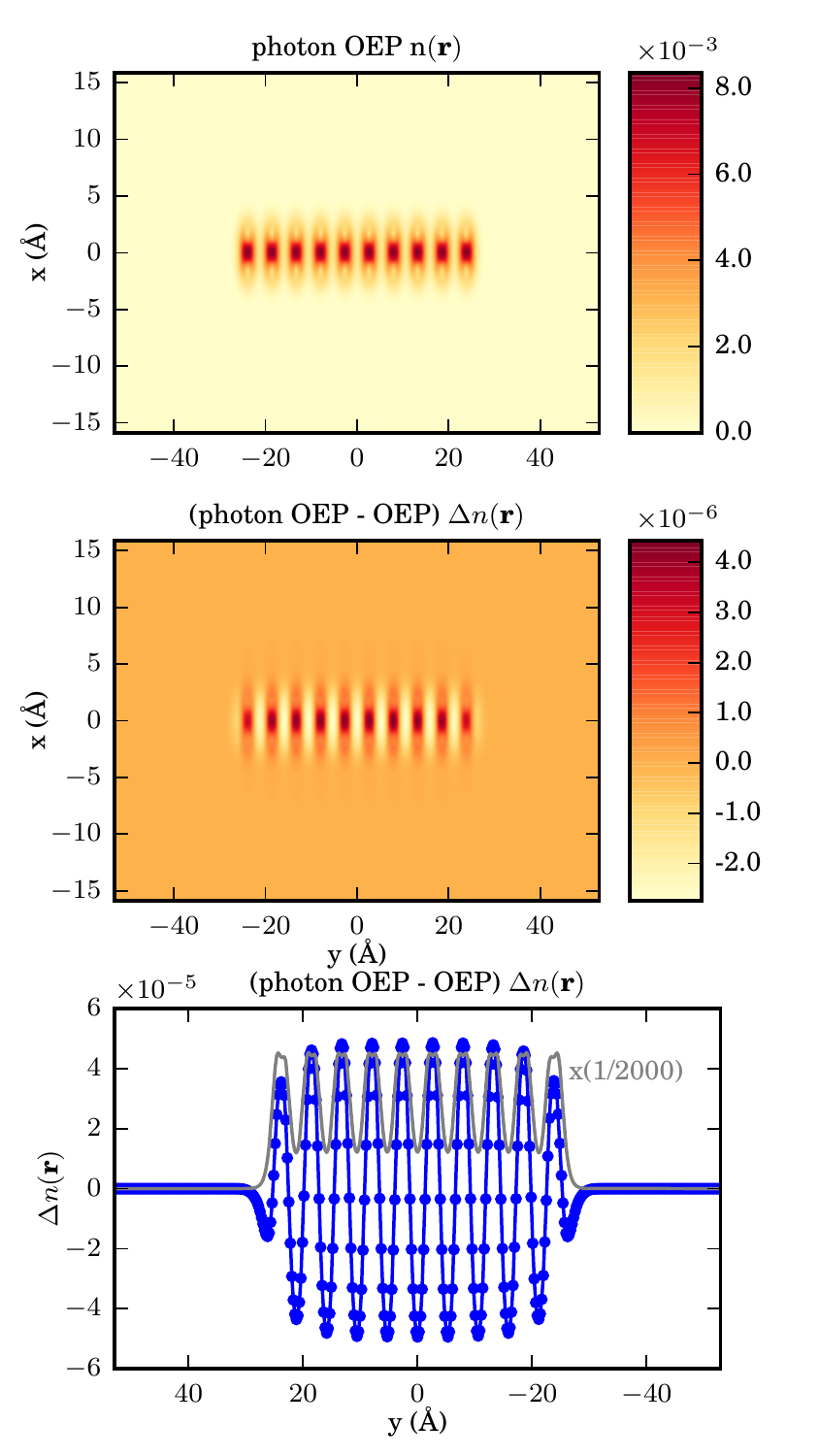}}
\caption{From top to bottom as a cut in x-y plane: OEP ground-state density of ten sodium dimers, difference of electron-photon OEP and electron OEP ground-state density, and cut of difference of electron-photon OEP and electron OEP along the y axis {in blue against the electron-photon OEP density in grey. Please note that the latter has been reduced by a factor of $1/2000$}.}
\label{fig:oep-rs-sodium-10}
\end{figure}
We compare to the KLI approximation in Fig.~\ref{fig:oep-rs-sodium-10-kli}. Here we find the KLI strongly overestimates the effects of the electron-photon interaction.
\begin{figure}[ht] 
\centerline{\includegraphics[width=0.5\textwidth]{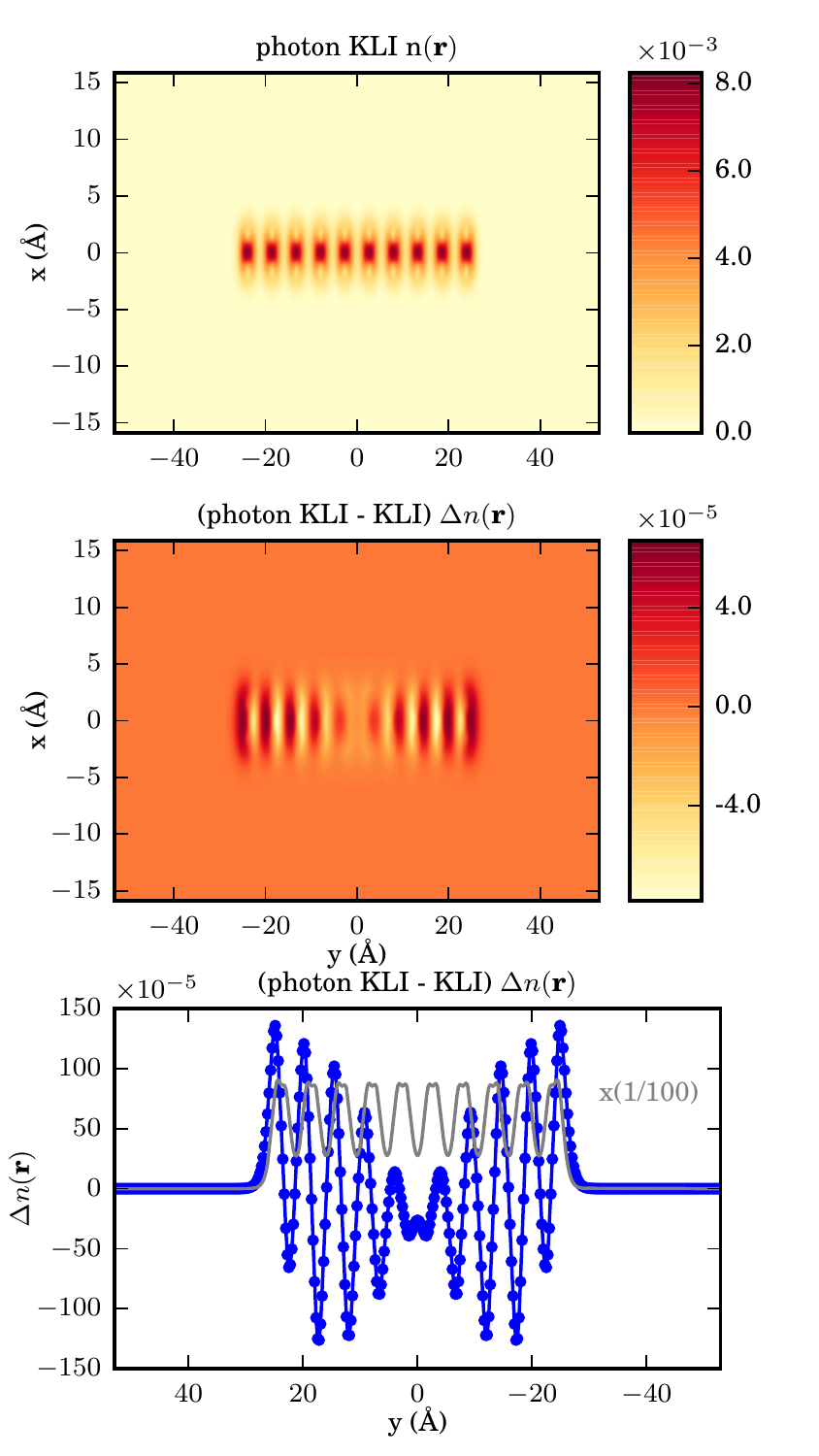}}
\caption{From top to bottom as a cut in x-y plane: KLI ground-state density of ten sodium dimers, difference of electron-photon KLI and electron KLI ground-state density, and cut of difference of electron-photon KLI and electron KLI along the y axis {in blue against the electron-photon OEP density in grey. Please note that the latter has been reduced by a factor of $1/100$}.}
\label{fig:oep-rs-sodium-10-kli}
\end{figure}
\begin{figure}[ht]
\centerline{\includegraphics[width=0.5\textwidth]{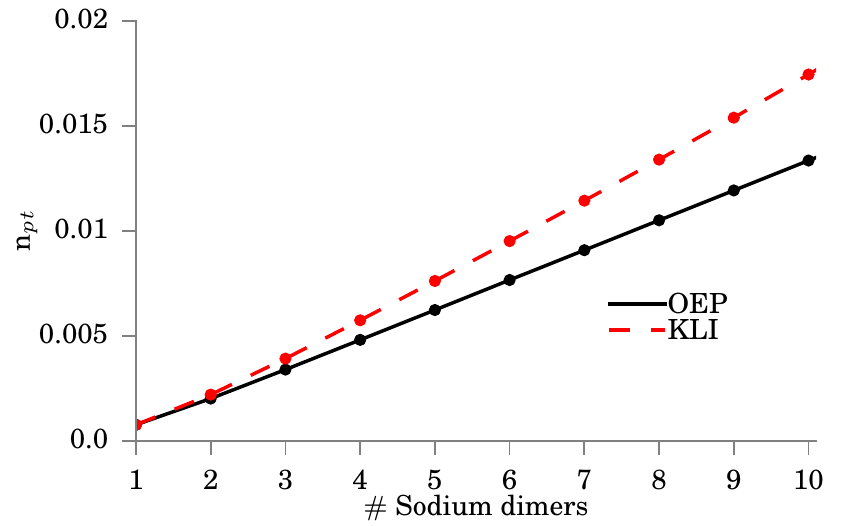}}
\caption{Photon occupation for the case of variable chain length of sodium dimers.}
\label{fig:oep-rs-sodium-npt}
\end{figure}
In the last figure of this section, Fig.~\ref{fig:oep-rs-sodium-npt}. We plot the number of photons in the correlated electron-photon ground state using the functional presented in Eq.~\ref{eq:ada}. In total, we find for the KLI and the OEP a linear behavior, where the KLI overestimates the number of photons slightly. {From Eq.~\ref{eq:ada} we also see that $\langle \hat{a}^\dagger_\alpha  \hat{a}_{\alpha} \rangle \sim \lambda_\alpha^2$.} Thus, we can {capture this behavior alternatively by} defining a new effective coupling constant $\tilde{\lambda}_\alpha \sim \sqrt{N_c}$ that scales with the square-root of the chain length. This example nicely illustrates the collective coupling of matter to the cavity field in the weak-coupling regime. This result agrees with predications based on the Dicke model, where the coupling strength scales with the square root of the number of two-level systems. However still more work needs to be done to properly characterize the emergence of collective phenomena due to the strong light-matter coupling in a set of $N$ emitters.

\section{Summary and Conclusion}
In conclusion, in this work, we have illustrated how the cavity-mediated electron-photon interaction is capable of rearranging the electron density in two- and three-dimensional systems. We find that our OEP approach accurately describes situations in the weak coupling limit. In the strong coupling limit, we find broken symmetry solutions which can be attributed to the accuracy of the employed approximate transversal energy orbital functional. The overall effect of the transversal photons on the ground state density is minor as expected from the magnitude of the Lamb-shift-type-energy correction. {However, it allows to investigate problems of quantum optics from first principles, such as the increase of the effective matter-photon coupling strength upon increasing the number of molecules inside a cavity. Furthermore,} the present work lays the foundation for the ab-initio construction of excited states and new functionals for QEDFT. While the small contribution of transversal photons on the ground state is reaffirming standard DFT calculations that neglect coupling to transversal photons, we have found that the effect of transversal photons on excited states, such as e.g. Rabi splittings, etc. can be substantial. The present work constitutes the first mandatory step towards such studies of excited states of strong light-matter coupled quantum systems. Additionally, this approach could also benefit standard electronic DFT, since similar ideas, i.e. expressing the exchange-correlation energy in terms of orbital shifts could also be applied to the correlation part in the xc approximation.\\
We have introduced our QEDFT approach as a viable tool to predict and describe the emerging field of QED chemistry, where chemical systems are placed in optical cavities.

\section{Acknowledgements}
We would like to thank Claudiu Genes, Camilla Pellegrini, and Ilya V. Tokatly for insightful discussions and acknowledge financial support from the European Research Council (ERC-2015-AdG-694097), by the European Union's H2020 program under GA no.676580 (NOMAD), and the Austrian Science {Fund} (FWF P25739-N27).

\section{Appendix}

\subsection{Derivation of the functional derivative of second orbital shift}
\label{sec:derivation_second_orbital_shift}
The derivation of the functional derivative of the second orbital shift with respect to the Kohn-Sham potential $v_s$ can be derived analogously to the derivation of Eq.~\ref{eq:wavefunction-functional} discussed in Ref.~\cite{engel2011}. By keeping the first order terms, we find the following Sternheimer equation that defines the infinitesimal change in the orbital shift
\begin{align}
&\left(\epsilon_{i\sigma} - \hat{T} - v_{s\sigma}(\textbf{r})  - \omega_\alpha\right) {\delta \Phi^{(1)}_{i\sigma,\alpha}(\textbf{r})} =  \left(\delta v_{s\sigma}(\textbf{r})-\delta \epsilon_{i\sigma} \right){\Phi^{(1)}_{i\sigma,\alpha}}(\textbf{r}) \nonumber\\
&+  \sqrt{\frac{\omega_\alpha}{2}}\hat{d}_\alpha { \delta \varphi_{i\sigma} (\textbf{r})}-\sqrt{\frac{\omega_\alpha}{2}} \sum_{k=1}^{N_\sigma}\Bigg[\bra{\varphi_{k\sigma} }\hat{d}_\alpha \ket{\varphi_{i\sigma}  } { \delta \varphi_{k\sigma}(\textbf{r}) }\nonumber\\
&+\bra{ \delta \varphi_{k\sigma} }\hat{d}_\alpha \ket{\varphi_{i\sigma}  } {\varphi_{k\sigma}(\textbf{r})  }+\bra{\varphi_{k\sigma} }\hat{d}_\alpha \ket{\delta \varphi_{i\sigma} } {\varphi_{k\sigma} (\textbf{r}) }\Bigg].
\end{align}
For $\delta \epsilon_{i\sigma}$, we can employ the following relation~\cite{engel2011}
\begin{align}
\delta \epsilon_{i\sigma} = \int d^3 r \varphi^*_{i\sigma}(\textbf{r}) \delta v_s(\textbf{r}) \varphi_{i\sigma}(\textbf{r})
\end{align}
The Sternheimer equation can be solved explicitly and has the solution
\begin{align}
\label{eq:delphi1}
\delta \Phi^{(1)}_{i\sigma,\alpha}(\textbf{r}) &= \int d^3 r' L^{(\alpha)}_{i\sigma}(\textbf{r},\textbf{r}') \left(\delta v_s(\textbf{r}') - \bra{\varphi_{i\sigma}}\delta v_s\ket{\varphi_{i\sigma}}\right)\Phi^{(1)}_{i\sigma,\alpha}(\textbf{r}') \\
&+ \sqrt{\frac{\omega_\alpha}{2}}K^{(\alpha)}_{i\sigma}(\textbf{r},\textbf{r}') \hat{d}_\alpha\delta \varphi_{i\sigma}(\textbf{r}')\nonumber\\
&-\sqrt{\frac{\omega_\alpha}{2}}\sum_{k=1}^{N_\sigma} \Bigg[\bra{\varphi_{k\sigma} }\hat{d}_\alpha \ket{\varphi_{i\sigma}  }  L^{(\alpha)}_{i\sigma}(\textbf{r},\textbf{r}'){ \delta \varphi_{k\sigma}(\textbf{r}') } \nonumber\\
&+\bra{ \delta \varphi_{k\sigma} }\hat{d}_\alpha\ket{\varphi_{i\sigma}  } L^{(\alpha)}_{i\sigma}(\textbf{r},\textbf{r}'){\varphi_{k\sigma} (\textbf{r}') }\Bigg],\nonumber
\end{align}
with the Greens functions
\begin{align}
K^{(\alpha)}_{i\sigma}(\textbf{r},\textbf{r}') = \sum_{l=N_\sigma+1}^{\infty} \frac{\varphi_{l\sigma}(\textbf{r})\varphi^*_{l\sigma}(\textbf{r}')}{\epsilon_{i\sigma} - \epsilon_{l\sigma}-\omega_\alpha},\\
L^{(\alpha)}_{i\sigma}(\textbf{r},\textbf{r}') = \lim_{\nu\rightarrow 0}\sum_{l=1}^{\infty} \frac{\varphi_{l\sigma}(\textbf{r})\varphi^*_{l\sigma}(\textbf{r}')}{\epsilon_{i\sigma} - \epsilon_{l\sigma}-\omega_\alpha + i \nu}.
\end{align}
By using Eq.~\ref{eq:delphi1}, we can calculate $\delta \Phi^{(1)}_{i\sigma,\alpha}(\textbf{r})/\delta v_{s\sigma}(\textbf{r}')$ and find Eq.~\ref{eq:second-oder-derivative}.

\bibliography{01_light_matter_coupling} 

\end{document}